\documentclass[conference,compsoc]{IEEEtran}
\usepackage{cite}
\usepackage{url}
\usepackage{xspace}
\usepackage{booktabs}
\usepackage{amsmath,amssymb,amsfonts}
\usepackage{algorithmic}
\usepackage{graphicx}
\usepackage{subcaption}
\usepackage{textcomp}
\usepackage{xcolor}
\usepackage{hyperref}
\usepackage{hhline}
\usepackage{float}
\usepackage{svg}
\PassOptionsToPackage{hyphens}{url}
\usepackage{pifont}
\usepackage{spverbatim}
\usepackage{listings}   
\usepackage{caption}    
\usepackage{multirow}
\usepackage{array}
\usepackage{wasysym}
\usepackage{tabularx,soul,mdframed,multirow}
\usepackage{algorithm2e}
\RestyleAlgo{ruled}
\usepackage{graphicx}
\usepackage{color}
\definecolor{lightgray}{rgb}{.9,.9,.9}
\definecolor{darkgray}{rgb}{.4,.4,.4}
\definecolor{purple}{rgb}{0.65, 0.12, 0.82}
\usepackage{stfloats}
\usepackage[table]{xcolor}

\usepackage{graphicx}
\usepackage{subcaption}
\usepackage{tikz}
\usepackage{setspace}
\usepackage{graphicx}
\usepackage{makecell}

\newcommand{\paragraphtitle}[1]{\vspace{5pt}\noindent\textbf{#1.}}

\newenvironment{icompact}{
  \begin{list}{$\bullet$}{
    \parsep 0pt plus 1pt            
    \partopsep 0pt plus 1pt         
    \topsep 2pt plus 2pt minus 1pt  
    \itemsep 4pt plus 1pt           
    \parskip 2pt plus 1pt           
    \leftmargin 0.13in              
    \labelwidth 0.13in
    }}
  {\normalsize\end{list}}

\newcounter{circlednum}

\newcommand*\circled[1]{\tikz[baseline=(char.base)]{
            \node[shape=circle,fill,inner sep=0.5pt] (char) {\tiny \textcolor{white}{#1}};}}

\newcommand{\twodigitcircled}{%
  \ifnum\value{circlednum}<10 0\fi\arabic{circlednum}%
}

\definecolor{sunnyyellow}{rgb}{1.00, 0.85, 0.10}      
\definecolor{turquoise}{rgb}{0.00, 0.75, 0.80}  
\definecolor{raspberry}{rgb}{0.8, 0.255, 0.396}  

\definecolor{roseblush}{rgb}{0.894, 0.8, 0.859}  

\definecolor{overleaforange}{rgb}{0.95, 0.61, 0.07}   
\definecolor{electricblue}{rgb}{0.12, 0.56, 1.00}     
\definecolor{springgreen}{rgb}{0.00, 0.80, 0.45}      
\definecolor{brightviolet}{rgb}{0.72, 0.35, 0.98}     
\definecolor{charcoalgray}{rgb}{0.30, 0.33, 0.38}     

\definecolor{oceanblue}{rgb}{0.27, 0.53, 0.84}     
\definecolor{leafgreen}{rgb}{0.35, 0.75, 0.49}     
\definecolor{deepviolet}{rgb}{0.58, 0.44, 0.86}    
\definecolor{coolgray}{rgb}{0.55, 0.57, 0.67}      
\definecolor{babyblue}{rgb}{0.788, 0.855, 0.973}
\definecolor{lavender}{rgb}{0.85, 0.82, 0.91}
\definecolor{green}{HTML}{008000}
\definecolor{lightgreen}{rgb}{0.41, 0.66, 0.31}        
\definecolor{darkgray}{rgb}{0.4, 0.4, 0.4}        
\definecolor{purple}{rgb}{0.6, 0, 1}
\definecolor{pink}{rgb}{0.73, 0, 0.39}         
\definecolor{red}{rgb}{0.7, 0, 0}                 
\definecolor{blue}{rgb}{0, 0, 1}                  
\definecolor{black}{rgb}{0, 0, 0}                 
\definecolor{regexcolor}{rgb}{0.73, 0.4, 0.53}    
\definecolor{builtincolor}{rgb}{0.90, 0.57, 0.22}        
\definecolor{stringcolor}{rgb}{0.61, 0.27, 0.22}  
\definecolor{pinkbg}{rgb}{1.0, 0.92, 0.95}

\definecolor{keywordcolor}{HTML}{008000}

\lstdefinelanguage{Python}{
  keywords={
    def, return, if, elif, else, for, while, in, is, not, and, or, 
    import, from, as, class, True, False, None, pass, break, continue, dict, for, in, global
  },
  keywordstyle=\color{green}\bfseries,
  morekeywords=[2]{len, split, enumerate, dict, isinstance, hasattr, setattr, getattr, get},
  keywordstyle=[2]\color{green}\ttfamily,
  sensitive=true,
  comment=[l]{\#},
  morecomment=[s]{/*}{*/},
  commentstyle=\color{purple}\ttfamily,
  stringstyle=\color{pink}\ttfamily,
  morestring=[b]",  
  morestring=[b]',  
  identifierstyle=\color{black},
  emph={set_property_value, @require_POST, message, ComponentRequest, View, create, handle, @classmethod, foo},
  emphstyle=\color{blue},
  escapechar=ß  
}

\usepackage{listings}
\usepackage{xcolor}

\definecolor{codegreen}{rgb}{0,0.6,0}
\definecolor{codegray}{rgb}{0.5,0.5,0.5}
\definecolor{codepurple}{rgb}{0.58,0,0.82}
\definecolor{codeblue}{rgb}{0.13,0.29,0.53}
\definecolor{codebg}{rgb}{0.95,0.95,0.95}

\lstdefinelanguage{JavaScript}{
  keywords={typeof, new, true, false, catch, function, return, null, catch, switch, var, if, in, while, do, else, case, break, const, let, async, await, export, import, class, extends, super, this},
  keywordstyle=\color{codeblue}\bfseries,
  ndkeywords={boolean, throw, import, export, typeof, document, window, console, require, module, process},
  ndkeywordstyle=\color{codepurple}\bfseries,
  identifierstyle=\color{black},
  sensitive=true,
  comment=[l]{//},
  morecomment=[s]{/*}{*/},
  commentstyle=\color{codegreen}\ttfamily\itshape,
  stringstyle=\color{red}\ttfamily,
  morestring=[b]',
  morestring=[b]",
  morestring=[b]`
}

\newcommand{\SecJS}{\textsc{SecJS}\xspace}
\newcommand{\syseval}{\textsc{JudgeJS}\xspace}

\newcommand{\benchgen}{\textsc{ForgeJS}\xspace} 

\newcommand{\benchname}{\textsc{ArenaJS}\xspace}

\usepackage{booktabs}
\usepackage{multirow}
\usepackage{makecell}
\usepackage{tabularx}
\usepackage{amssymb}
\usepackage[figuresright]{rotating}

\makeatletter
\newcommand{\linebreakand}{%
  \end{@IEEEauthorhalign}
  \hfill\mbox{}\par
  \mbox{}\hfill\begin{@IEEEauthorhalign}
}
\makeatother

\usetikzlibrary{chains,shadows.blur,positioning,arrows.meta}
\begin{document}

\title{Large Language Models Cannot Reliably Detect Vulnerabilities in JavaScript: \\
The First Systematic Benchmark and Evaluation}
\IEEEoverridecommandlockouts
 
\author{
\IEEEauthorblockN{Qingyuan Fei}
\IEEEauthorblockA{Lanzhou University\\
feiqy2023@lzu.edu.cn}
\and
\IEEEauthorblockN{Xin Liu}
\IEEEauthorblockA{Lanzhou University\\
bird@lzu.edu.cn}
\and
\IEEEauthorblockN{Song Li}
\IEEEauthorblockA{Zhejiang University\\
songl@zju.edu.cn}
\and
\IEEEauthorblockN{Shujiang Wu}
\IEEEauthorblockA{Beihang University\\
wushujiang@buaa.edu.cn}
\linebreakand
\IEEEauthorblockN{Jianwei Hou}
\IEEEauthorblockA{Data and Technology Support Center\\
of the Cyberspace Administration of China\\
houjianwei@cac.gov.cn}
\and
\IEEEauthorblockN{Ping Chen}
\IEEEauthorblockA{Fudan University\\
pchen@fudan.edu.cn}
\and
\IEEEauthorblockN{Zifeng Kang}
\IEEEauthorblockA{Beijing University of\\
Posts and Telecommunications\\
zifengkang@bupt.edu.cn}
\thanks{Corresponding authors: Zifeng Kang and Xin Liu.}
}



\maketitle  
\pagestyle{plain}

\begin{abstract}
    Researchers have proposed numerous methods to detect vulnerabilities in JavaScript, especially those assisted by Large Language Models (LLMs).
    However, the actual capability of LLMs in JavaScript vulnerability detection remains questionable, necessitating systematic evaluation and comprehensive benchmarks. 
    Unfortunately, existing benchmarks suffer from three critical limitations: (1) incomplete coverage, such as covering a limited subset of CWE types, (2) underestimation of LLM capabilities caused by unreasonable ground truth labeling, and (3) overestimation due to unrealistic cases such as using isolated vulnerable files rather than complete projects.
    
    In this paper, we introduce, for the first time, three principles for constructing a benchmark for JavaScript vulnerability detection that directly address these limitations: (1) comprehensiveness, (2) no underestimation, and (3) no overestimation.
    Guided by these principles, we propose \benchgen, the first automatic benchmark generation framework for evaluating LLMs' capability in JavaScript vulnerability detection. 
    Then, we use \benchgen to construct \benchname---the first systematic benchmark for LLM-based JavaScript vulnerability detection---and further propose \syseval, an automatic evaluation framework. 
    
    We conduct the first systematic evaluation of LLMs for JavaScript vulnerability detection, leveraging \syseval to assess seven popular commercial LLMs on \benchname. 
    The results show that LLMs not only exhibit limited reasoning capabilities, but also suffer from severe robustness defects, indicating that reliable JavaScript vulnerability detection with LLMs remains an open challenge.
\end{abstract}

\section{Introduction}
\label{sec:introduction}

JavaScript is the core language most widely deployed in browsers and increasingly widespread across server-side ecosystems; its security directly impacts user data protection and the software supply chain. In recent years, large language models (LLMs) have demonstrated strong general capabilities in code understanding, generation, and auditing, motivating their adoption for JavaScript vulnerability detection. In parallel, classic program analysis for JavaScript security has progressed substantially\cite{li2022mining,li2021detecting,kang2023scaling,kang2022probe,liu2024undefined}. However, evidence from practice suggests that the true effectiveness of this direction remains highly questionable: on real projects, models exhibit persistently high false positive rates, detection accuracy is difficult to reproduce stably, and behavior is highly sensitive to contextual shifts and code perturbations, revealing pronounced unreliability. More fundamentally, the community still lacks a systematic benchmark that comprehensively covers security risks in the JavaScript ecosystem, mitigates evaluation bias, and enables automated reproducibility, leading to contradictory conclusions across settings and hindering practical deployment.

To address these challenges, we propose three guiding principles for constructing a benchmark for JavaScript vulnerability detection: comprehensiveness, no underestimation, and no overestimation. Comprehensiveness demands broad coverage of weakness types, evaluation dimensions, and data sources to reduce evaluation bias and improve generalizability across diverse settings. No underestimation emphasizes mechanisms such as semantic equivalence classes and fuzzy matching to accommodate variability in model outputs and uncertainty in real-world annotations, preventing excessive penalties from rigid string matching or noisy labels. No overestimation requires industry-realistic, project-level evaluation that explicitly quantifies the cost of false positives and probes reliance on superficial cues via diverse data-augmentation strategies, thus suppressing artificially inflated laboratory results. Taken together, these principles aim to objectively define the realistic upper and lower bounds of LLM performance under reproducible, scalable, deployment-constrained conditions.

Guided by these principles, we introduce \benchgen, the first automated generation framework for constructing LLM-based benchmarks for JavaScript vulnerability detection. To achieve comprehensiveness, \benchgen aggregates heterogeneous sources, covers 218 CWE types, unifies extraction and normalization pipelines, and supports both function-level and project-level evaluation to align weakness distributions and contextual complexity with real-world characteristics. To avoid underestimation, \benchgen relaxes brittle exact string alignment via CWE equivalence classes and fuzzy matching, introduces a dual-track evaluation with complete versus denoised datasets to isolate annotation noise, and calibrates benchmark labels against state-of-the-art reference frameworks (e.g., claude-code-security-review). To avoid overestimation, \benchgen uses complete projects rather than single-file snippets, constructs pre- and post-fix project pairs to directly quantify false positives and localize error sources, and systematically introduces four augmentation strategies to disrupt reliance on filenames, import paths, and comments.

Building on \benchgen, we construct \benchname, the first systematic benchmark for LLM-based JavaScript vulnerability detection, spanning multiple granularities, combining real-world and synthetic data, and unifying multi-perspective metrics. To enable large-scale, reproducible comparisons, we propose \syseval, an automated evaluation framework that spans prompt templates, response parsing, label alignment, and robust scoring; leveraging semantic equivalence and fuzzy matching, it harmonizes heterogeneous outputs and yields a unified, rigorous, and traceable metric suite across function- and project-levels, including F1, false positive rate (FPR), and detection efficacy under engineering constraints.

Using \benchname and \syseval, we systematically evaluate seven mainstream commercial LLMs, including GPT-5. The results reveal three findings of substantive impact. First, models exhibit missing reasoning pathways: at the project level, F1 surpasses the function level by 8\%--18\%, but the gain is driven largely by surface features (filenames, imports, comments) rather than taint propagation and dependency analysis, indicating reliance on opportunistic heuristics. Second, robustness is deficient: seemingly mild augmentations and noise induce drastic swings—for example, under specific noise conditions, a leading model's F1 drops from 35.9\% to 4.2\%—and compounded augmentations can trigger catastrophic failures, evidencing high sensitivity to fragile cues. Third, from the perspective of the VD-S metric~\cite{ding2024primevul}, we find that models are difficult to deploy in realistic industrial settings: with $\mathrm{FPR}\le 0.5\%$, the system misses more than three-quarters of real vulnerabilities and cannot sustain acceptable detection performance.

In summary, our contributions are threefold.

\begin{icompact}
    \item We establish a systematic methodology for JavaScript vulnerability benchmarking and operationalize three principles—comprehensiveness, no underestimation, and no overestimation.
    \item We deliver the first systematic benchmark and evaluation for LLM-based JavaScript vulnerability detection: the automated generation framework \benchgen, the broad-coverage benchmark \benchname, and the scalable, reproducible evaluation framework \syseval.
    \item We provide a systematic evaluation that reveals key limitations in reasoning sufficiency, robustness, and real-world usability and offers a reproducible testbed to drive progress toward reliable, deployable LLM-based vulnerability detection.
\end{icompact}

\section{Overview}
\label{sec:overview}
\subsection{Background: LLM for JavaScript Vulnerability Detection}
Large language models (LLMs) and deep learning approaches have recently been applied to vulnerability detection tasks, demonstrating capabilities across multiple programming languages\cite{li2018vuldeepecker,zhou2019devign,hanif2022vulberta,li2023vulnerability,wang2024extensive}. Unlike traditional static application security testing (SAST) tools\cite{jovanovic2006pixy,kashyap2014jsai,owasp2021top10} that rely on predefined rules and pattern matching, LLMs are designed to identify vulnerabilities through learned representations of code semantics and patterns.

Nevertheless, in JavaScript security analysis, current LLMs often over-rely on surface-level pattern matching rather than robust semantic understanding. This manifests in three ways: (1) keyword dependence; (2) syntactic-structure dependence; and (3) neglect of execution context and complete taint flows from sources to sinks.
Illustrative code examples are provided in the Appendix (Listings~\ref{lst:ov_keyword}, \ref{lst:ov_syntax}, \ref{lst:ov_context}).

Existing studies provide limited, systematic evaluation of LLMs' actual capability on JavaScript vulnerability detection. Many benchmarks such as SECBENCH.JS~\cite{secbenchjs} and VulcaN~\cite{shcherbakov2023silent} use code snippets instead of complete projects, which fails to reflect the complexity of real-world repositories\cite{kang2022probe,shcherbakov2023silent,li2021detecting}. Moreover, evaluation protocols commonly assess only whether vulnerability is detected and ignore reasoning quality such as whether the model can correctly localize the vulnerable file, function, and CWE type. Two extremes are prevalent: evaluations can overestimate capability by simplifying tasks via snippet inputs, or underestimate capability due to inconsistent labeling granularity when the model reports CWE-83 while the ground truth uses CWE-79 for an equivalent attack.

This problem is especially salient for JavaScript. As the language that runs both in browsers on the frontend and on servers via Node.js backend, vulnerability characteristics vary significantly by context—frontend issues often involve DOM-based XSS and client-side injections~\cite{owasp2021top10}, whereas backend issues include SQL injection, command injection, and prototype pollution\cite{nodejs,owasp2021top10}. Existing JavaScript datasets such as SECBENCH.JS include only a few hundred samples and primarily server-side cases, making comprehensive evaluation difficult. A benchmark that spans thousands of real JavaScript projects and simultaneously avoids both overestimation and underestimation remains lacking. 

\subsection{A Motivating Example}
We identify a real prototype pollution case from GitHub, specifically CVE-2021-25941 in project ASaiAnudeep/deep-override\cite{nvd:cve-2021-25941,deep-override:repo}, that exemplifies nine deficiencies in existing benchmarks and highlights large language models' tendency to overfit lexical patterns rather than semantics. The vulnerable project is a backend library for deep-merging JavaScript objects. The flaw resides in the function \texttt{override} in \texttt{src/index.js}.

\begin{lstlisting}[float=!t,caption={Code comparison (vulnerable vs. fixed version).},label={lst:motivating_vuln_fixed},language=JavaScript,numbers=left,basicstyle=\ttfamily\footnotesize]
const isObject = require('./utils/isObject');
const cloneDeep = require('./utils/cloneDeep');

function override(target, source, options = {}) {
  // ... parameter validation and circular reference handling ...

  // Main merge logic
  for (const key in source) {
    if (!source.hasOwnProperty(key)) continue;

    // VULNERABILITY: only blocks __proto__, misses constructor & prototype
-   if (key === '__proto__') continue; // Incomplete protection
+   // FIX: block all prototype pollution vectors
+   const dangerousKeys = ['__proto__', 'constructor', 'prototype'];
+   if (dangerousKeys.includes(key)) continue; // Complete protection

    const sourceValue = source[key];

    // Deep merge for nested objects
    if (isObject(sourceValue)) {
      if (isObject(target[key])) {
        target[key] = override(target[key], sourceValue, options);
      } else {
        target[key] = options.clone ? cloneDeep(sourceValue) : sourceValue;
      }
    } else {
      target[key] = sourceValue;
    }
  }

  // ... other option handling ...
  return target;
}

module.exports = override;
\end{lstlisting}

\begin{lstlisting}[float=!t,caption={Exploitation example.},label={lst:motivating_exploit},language=JavaScript,numbers=left,basicstyle=\ttfamily\footnotesize,breaklines=true]
// Exploit payload (via constructor.prototype)
const maliciousPayload = JSON.parse(
  '{"constructor":{"prototype":' +
  '{"isAdmin":true,"role":"admin"}}}'
);
// Trigger pollution
override({}, maliciousPayload);
// Global pollution takes effect
const normalUser = {};
console.log(normalUser.isAdmin);  // true - unintended admin privilege
console.log(normalUser.role);     // "admin"
// Affects subsequently created objects
const anotherUser = { name: "Alice" };
console.log(anotherUser.isAdmin); // true
\end{lstlisting}

\paragraphtitle{Defect 1: Incomplete CWE coverage} This vulnerability is prototype pollution (CWE-1321)\cite{mitre:cwe-1321}, a JavaScript-specific class formally added in 2019. Existing benchmarks emphasize traditional CWEs: SECBENCH.JS\cite{kang2022probe} includes only five CWE types while omitting many newer categories added after 2018 (e.g., CWE-611 XXE injection)\cite{mitre:cwe-611}, creating substantial evaluation gaps for LLM capability on emerging weakness detection.

\paragraphtitle{Defect 2: One-dimensional evaluation} Many evaluations only verify the binary outcome (vulnerability present or not) without requiring reasoning quality. For this case, if a model merely outputs ``prototype pollution vulnerability, severity: high,'' it would be deemed a correct detection. Yet the model may fail to answer critical questions: (1) Which file? (\texttt{src/index.js}, not \texttt{utils/isObject.js} or \texttt{test/test.js}); (2) Which function? (the \texttt{override} function); (3) Which exact line? (line 12's incomplete check, not line 8's \texttt{for...in} loop or line 27's assignment); (4) Why is line 12 inadequate? (blocks only \texttt{\_\_proto\_\_}, allowing \texttt{constructor} or \texttt{prototype} bypass). If the model answers via keyword matching (searching \texttt{\_\_proto\_\_} or \texttt{prototype}) without understanding taint flow (\texttt{source[key]} $\rightarrow$ \texttt{target[key]} assignment chain), such ``correct detection'' is mere surface pattern matching.

\paragraphtitle{Defect 3: Insufficient data comprehensiveness} This backend library, operating in the Node.js ecosystem and utilized by web frameworks including Express and Koa, exemplifies cases where existing datasets fail to stratify frontend, backend, and full-stack contexts. JavaScript vulnerability patterns vary substantially across these environments: backend prototype pollution may enable privilege escalation or remote code execution (via \texttt{user.isAdmin} or \texttt{child\_process} module pollution), while frontend prototype pollution predominantly causes DOM-based XSS or client-side denial of service. Mixed evaluation obscures LLM performance variations across distinct contexts.

Furthermore, the full repository comprises 15 files totaling 2,347 lines of code (\texttt{src/} directory: 5 files, 823 lines; \texttt{test/} directory: 7 files, 1,124 lines; \texttt{node\_modules/} directory: 400 lines), but existing evaluations extract only the vulnerable function (approximately 33 lines), entirely discarding contextual dependencies including the \texttt{isObject} utility implementation, legitimate usage patterns in test cases, and security-relevant \texttt{package.json} policies. For example, this project's \texttt{engines} field specifies \texttt{"node": ">=12.0.0"} to mandate Node.js 12 or later, thus circumventing \texttt{Object.prototype} pollution vulnerabilities known to affect earlier versions.

\paragraphtitle{Defect 4: Inconsistent CWE granularity} This vulnerability's official NVD label is CWE-1321 (Improperly Controlled Modification of Object Prototype Attributes `Prototype Pollution'), yet labeling granularity varies across sources: Mend.io assigns CWE-915 (Improperly Controlled Modification of Dynamically-Determined Object Attributes), Snyk uses CWE-471 (Modification of Assumed-Immutable Data), and GitHub Advisory uses CWE-1321\cite{github:advisory}. If an LLM reports ``CWE-915 vulnerability: attackers can dynamically modify object attributes for privilege escalation,'' the description is entirely accurate, yet strict CWE-ID matching would incorrectly penalize it as a type mismatch. This labeling granularity inconsistency is widespread—manual analysis of 50 JavaScript vulnerability samples reveals that 16\% exhibit CWE annotation discrepancies across sources (e.g., CWE-79 vs. CWE-83, CWE-89 vs. CWE-564). Strict matching causes model F1 scores on GPT-5 to drop from 0.3207 to 0.2681, severely understating capability.

\paragraphtitle{Defect 5: Label noise contaminates evaluation} This project's GitHub fix commit (SHA: 2aced176, 2021-05-14) not only patches the vulnerability but also includes: (1) dependency version upgrades for 3 packages (\texttt{package.json}: lodash 4.17.19$\rightarrow$4.17.21, mocha 8.2.1$\rightarrow$8.4.0, chai 4.2.0$\rightarrow$4.3.4); (2) test suite restructuring (\texttt{test/test.js}: 152$\rightarrow$203 lines, 31 new test cases); (3) README updates (Security Policy section, 42 new lines). Naively comparing all pre-/post-fix differences (5 files, 137 line changes) produces annotation errors: mislabeling \texttt{package.json} as a ``vulnerable file'' (actually just dependency upgrades), \texttt{test/test.js} as ``vulnerable code'' (actual flaw is in \texttt{override}), and \texttt{README.md} as a ``fix file'' (merely documentation). When a model correctly identifies \texttt{src/index.js} line 12 but omits \texttt{package.json}, it is wrongly penalized for ``missing vulnerable files'' (though \texttt{package.json} is not vulnerable).

\paragraphtitle{Defect 6: Prompting underutilizes model capability} With overly simple prompts (e.g., ``Please analyze whether this code has security vulnerabilities''), LLMs may fail to leverage their full reasoning capabilities. This case demands complex multi-step analysis: (1) understanding JavaScript prototype chain semantics—recognizing \texttt{for...in} loops traverse both own and prototype-chain properties, and the relationships among \texttt{\_\_proto\_\_}, \texttt{constructor.prototype}, and \texttt{Object.prototype}; (2) tracing taint propagation paths—identifying the taint source as \texttt{source[key]} (line 17, attacker-controlled input), following dataflow through the \texttt{sourceValue} variable (line 17), conditional branches (line 20), recursive calls (line 22) or direct assignments (line 27), to the taint sink \texttt{target[key]} (lines 22 or 27, writing to the target object); (3) reasoning about attack vectors—understanding that when \texttt{key} is \texttt{constructor}, \texttt{target['constructor']} overwrites the constructor reference, enabling pollution of all instances via \texttt{constructor.prototype}; (4) assessing mitigation completeness—recognizing that line 12's \texttt{if (key === '\_\_proto\_\_')} blocks only one pollution vector, missing \texttt{constructor} and \texttt{prototype}. Simple prompts cannot guide models through these four reasoning steps, severely underestimating LLM capability.

\paragraphtitle{Defect 7: Snippet-only inputs reduce task realism} Extracting only the vulnerable function snippet (approximately 33 lines) allows models to identify \texttt{\_\_proto\_\_} (appearing once, line 12) via keyword matching without understanding: (1) project dependency relationships—this library is depended upon by 1,200+ npm packages (including \texttt{express-session}, \texttt{body-parser}, \texttt{config}, etc.), with the vulnerability affecting millions of Node.js applications; (2) call chain complexity—the attack path spans HTTP request $\rightarrow$ \texttt{body-parser} JSON parsing $\rightarrow$ \texttt{express-session} reading session data $\rightarrow$ \texttt{override} merging configuration $\rightarrow$ global object pollution $\rightarrow$ privilege check logic impact, requiring understanding of a 5-layer call stack; (3) contextual dependencies—the \texttt{isObject} utility function (\texttt{utils/isObject.js}) implementation affects line 20's conditional outcome; if \texttt{isObject} incorrectly classifies \texttt{null} as an object, additional vulnerability paths emerge; (4) test coverage—\texttt{test/test.js} contains 152 test cases, yet none test malicious inputs (e.g., objects with \texttt{constructor} properties), indicating developers were unaware of this threat vector. These contextual details are critical for assessing real-world vulnerability impact, but snippet-only inputs discard them entirely.

\paragraphtitle{Defect 8: No false-positive assessment} Existing evaluations provide only vulnerable-version code without validating model behavior on fixed versions. This prevents identifying keyword over-reliance—if the model's detection logic is ``search for \texttt{\_\_proto\_\_} keyword; if found, report prototype pollution,'' then even in the fixed code (lines 13--15 add complete protection: \texttt{const dangerousKeys = ['\_\_proto\_\_', 'constructor', 'prototype']; if (dangerousKeys.includes(key)) continue;}, which is actually secure), the model will still misreport a vulnerability upon seeing the \texttt{\_\_proto\_\_} string at line 14. Additionally, the fixed version presents another test: line 14's \texttt{dangerousKeys} array is defined inside the loop (recreated each iteration)—a performance issue, not a security flaw; models flagging this as a vulnerability commit false positives.

\paragraphtitle{Defect 9: Missing robustness testing} The pattern-matching over-reliance observed in Defect 8 becomes even more pronounced under noise and obfuscation. Applying four adversarial transformations to this vulnerable code reveals sharp drops in LLM detection rates, confirming reliance on surface features rather than semantic understanding:
\begin{icompact}
    \item \textbf{(a) Obfuscation.} After applying javascript-obfuscator\cite{js:obfuscator}, variable names like \texttt{key} become \texttt{\_0x3a2f}, and \texttt{\_\_proto\_\_} is encoded as Unicode escape \texttt{\textbackslash x5f\textbackslash x5f\textbackslash x70\textbackslash x72\textbackslash x6f\textbackslash x74\textbackslash x6f\textbackslash x5f\textbackslash x5f}. Models relying on clear identifiers and literal keywords will fail to recognize the vulnerability after obfuscation.

    \item \textbf{(b) Noise injection.} We insert 51 safe decoy sinks at function entry (e.g., \texttt{fs.appendFileSync} with hardcoded log paths, \texttt{db.execute} with hardcoded queries, \texttt{innerHTML} with hardcoded content)—seemingly dangerous yet safe (no taint sources). Models using naive ``flag all dangerous APIs'' heuristics will produce abundant false positives while missing the real vulnerability.

    \item \textbf{(c) Noise + obfuscation.} Both code obfuscation and noise injection are applied simultaneously. Models with genuine dataflow understanding should still localize the real vulnerability and suppress false positives under these interferences.

    \item \textbf{(d) Prompt injection.} Misleading comments are inserted before vulnerable code (e.g., \texttt{// SECURITY AUDIT PASSED}, \texttt{// All known attack vectors handled}, \texttt{// NO VULNERABILITIES FOUND}); models over-relying on comment text rather than logic analysis will misclassify vulnerable code as safe.
\end{icompact}

Collectively, these nine defects indicate that current LLMs do not truly understand security semantics in JavaScript vulnerability detection, but instead rely on surface features—clear variable names (e.g., \texttt{override}, \texttt{\_\_proto\_\_}), specific code patterns, and comment text cues. Such pattern dependence cannot handle code obfuscation, misleading documentation, and complex contexts in real-world scenarios, resulting in low detection rates.

\subsection{Limitations of Existing Benchmarks}
We conclude from the motivating example that existing benchmarks suffer from three core issues: \textbf{(1) Unrealistic data and task setups}—code snippets replace full projects, and frontend/backend/full-stack stratification is absent, failing to reflect real-world engineering complexity; \textbf{(2) Insufficient evaluation and annotation}—binary detection is emphasized over reasoning and localization quality, CWE granularity labeling is inconsistent and annotation quality is poor, undermining result comparability and credibility; \textbf{(3) Missing protocols and robustness testing}—prompting is overly simplistic, vulnerable/fixed project pairs for false-positive measurement are lacking, and adversarial tests (obfuscation, noise, prompt injection) are absent. A comprehensive, rigorous, and robust evaluation framework is therefore essential to accurately assess LLM vulnerability detection capability.

\subsection{Method Overview}
We propose \SecJS, a comprehensive benchmark designed to evaluate LLMs' capability in JavaScript vulnerability detection. \SecJS{} comprises a dataset generation framework (\benchgen{}) and an automated evaluation framework (\syseval{}). Its design is guided by three principles, each instantiated by three concrete techniques.

\paragraphtitle{Principle I: Comprehensiveness}
\begin{icompact}
  \item \textbf{Principle I-1: Comprehensive CWE Coverage.} \benchgen{} collects real projects across a broad range of CWEs, covering both traditional weaknesses, e.g., SQL injection, XSS, command injection, authentication bypass and hardcoded credentials; and newer categories, e.g., prototype pollution, ReDoS, and XXE. 
  \item \textbf{Principle I-2: Comprehensive Evaluation Methodology.} We evaluate not only detection outcomes but also reasoning and localization quality from both project- and function-level, requiring models to identify the CWE type as well as the vulnerable file and function locations when applicable. 
  \item \textbf{Principle I-3: Comprehensive Data Sources.} All samples are drawn from real GitHub repositories and stratified by application context---including frontend, backend and full-stack---to reflect ecosystem diversity\cite{githubCommunity}.
\end{icompact}

\paragraphtitle{Principle II: No underestimation}
\begin{icompact}
  \item \textbf{Principle II-1: Employing CWE Equivalence Classes.} We group related CWEs into equivalence classes so that mismatches in granularity (e.g., CWE-79 vs. CWE-83) do not unfairly bias the evaluation~\cite{capec}. 
  \item \textbf{Principle II-2: Denoising Datasets.} We provide noise-reduced subsets to mitigate annotation errors from unrelated edits (e.g., dependency upgrades, test refactors).
  \item \textbf{Principle II-3: Maximizing Detection Capability for LLMs.} We employ a strong prompting framework, claude-code-security-review~\cite{anthropic}, with state-of-the-art LLMs to encourage multi-step reasoning, including cataloging security patterns, contrasting code against known safe patterns, and tracing taint flows from sources to sinks.
\end{icompact}

\paragraphtitle{Principle III: No overestimation}
\begin{icompact}
  \item \textbf{Principle III-1: Repository-Level Context.} Models analyze full repositories instead of isolated code snippets, operating under realistic complexity and context.
  \item \textbf{Principle III-2: Vulnerable--Fixed Project Pairs.} Each case includes pre-fix and post-fix versions to quantify false positives and measure understanding of fixes.
  \item \textbf{Principle III-3: Enough Dataset Variants for Robustness Evaluation.} We construct noise, obfuscation, noise+obfuscation, and prompt-injection variants to probe robustness and detect reliance on superficial cues\cite{js:obfuscator}.
\end{icompact}

\providecommand{\paragraphtitle}[1]{\textbf{#1}}

\section{Methodology}
\label{sec:methodology}

In this section, we describe the architecture of \SecJS{} Framework, which comprises two key components: \benchgen{} (dataset generation framework) and the \syseval{} evaluation framework.

\subsection{System Architecture}
\begin{figure*}[t]
\centering
\includegraphics[width=\textwidth]{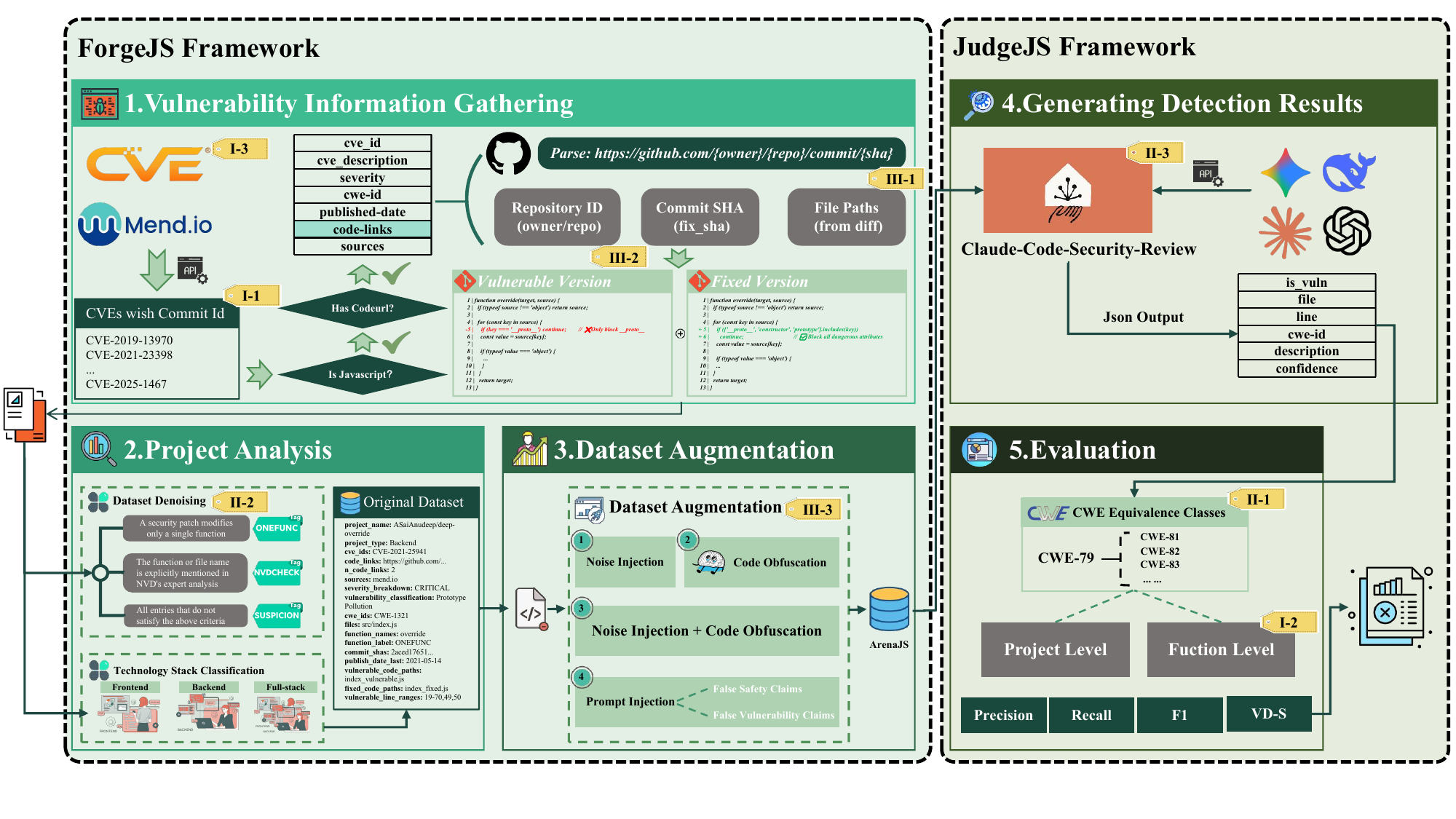}
\caption{Overall architecture of the \SecJS{} Framework.}
\label{fig:secjs-framework}
\end{figure*}

Figure~\ref{fig:secjs-framework} presents the overall architecture of the \SecJS{} Framework. The \SecJS{} Framework comprises two major components: (a) the \benchgen{} automated dataset generation framework and (b) the \syseval{} automated evaluation framework. For component (a), \benchgen{} builds the \benchname{} benchmark through three sequential steps.

\paragraphtitle{(1) Vulnerability Information Gathering} \benchgen{} collects vulnerability metadata from the official CVE website and Mend.io, requiring two essential fields: Metadata (e.g., CVE identifier, vulnerability type, summary) and GitHub URL. It then parses linked patch commits to extract repository identifiers, commit SHAs, affected file paths, and retrieves both pre-patch (parent commit) and post-patch (current commit) code versions.

\paragraphtitle{(2) Project Analysis} This step further enriches the dataset by performing multi-dimensional categorization (e.g., frontend/backend/full-stack, complete/denoised datasets) and refining ground truth annotations (i.e., pinpointing vulnerable functions and their corresponding line ranges).

\paragraphtitle{(3) Dataset Augmentation} To more effectively assess the true capabilities of large language models beyond simple pattern matching, we augment the original dataset with four enhancement strategies: noise injection, obfuscation, combined noise and obfuscation, and prompt injection, resulting in five datasets.

For component (b), \syseval{} leverages claude-code-security-review (currently the state-of-the-art prompt system for vulnerability detection) to query large language models including ChatGPT-5 and Grok-4 for vulnerability detection across the five dataset variants generated from component (a) and collect their detection outputs. The evaluation is conducted at two granularities: project-level using binary tuple matching (vulnerability presence, CWE type correctness) and function-level using quaternary tuple matching (vulnerability presence, CWE type correctness, vulnerable file identification, vulnerable function identification), measuring each LLM's performance with metrics such as precision, recall, F1-score, and VD-S.

\subsection{\benchname{} Benchmark}

\benchname{} is constructed automatically through three sequential steps: Vulnerability Information Gathering, Project Analysis, and Dataset Augmentation.

\subsubsection{Vulnerability Information Gathering}

We employ web crawlers to gather JavaScript vulnerability information from the official CVE website~\cite{cve} and Mend.io~\cite{mend}. Mend.io (formerly WhiteSource) is a commercial platform specializing in Software Composition Analysis (SCA)~\cite{ponta2020detection} and software supply chain security~\cite{ladisa2023sok}. Many CVE entries lack corresponding GitHub repository URLs (or contain obsolete links), while Mend.io effectively complements the CVE database in this aspect. The collected data must include two essential fields: Metadata (containing CVE identifier, CVE description, severity level, CWE type, and publication date) and GitHub URL (used for subsequent project retrieval). A complete example data entry after this step is provided in Appendix~\ref{app:vulnerability-gathering-example}.


Within the gathering stage, after acquiring CVE metadata from the NVD API~\cite{nvd} and Mend vulnerability database~\cite{mend}, \benchgen{} extracts the actual vulnerable and patched code from GitHub commit history~\cite{githubCommunity}. This sub-process maps CVE entries to concrete code changes and retrieves complete pre-patch and post-patch code files. CVE descriptions typically include links to GitHub patch commits (e.g., \texttt{https://github.com/user/repo/commit/sha}). The system parses these URLs to extract critical information:

\begin{icompact}
\item \textbf{Repository identifier} (\texttt{owner/repo}): identifying the affected GitHub repository
\item \textbf{Commit SHA} (\texttt{commit\_sha}): pinpointing the specific patch commit
\item \textbf{File path} (\texttt{file\_path}): extracting affected files from commit diff information
\end{icompact}

To construct complete vulnerable-patched code pairs, the system retrieves two code versions: the Vulnerable Version (based on the parent commit SHA) and the Fixed Version (based on the current commit SHA).

\subsubsection{Project Analysis}

This step further enriches the dataset. Specific operations include (1) multi-dimensional dataset categorization and (2) ground truth refinement. In part (1), we categorize the dataset along two dimensions: frontend/backend/full-stack and complete/denoised datasets.

\paragraphtitle{Complete vs. Denoised Dataset} For complete/denoised dataset categorization, we adhere to the principle of ``not underestimating LLM capabilities''. Empirical analysis reveals that numerous real-world projects simultaneously update versions while patching vulnerabilities, introducing substantial noise into the pre-patch and post-patch code pairs obtained during the gathering stage, thereby compromising ground truth accuracy. To address this, we assign three types of labels to dataset entries:

\begin{icompact}
\item \textbf{ONEFUNC}: assigned when a security patch modifies only a single function
\item \textbf{NVDCHECK}: assigned when the function or file name is explicitly mentioned in NVD's expert analysis
\item \textbf{SUSPICION}: assigned to all entries that do not satisfy the above criteria
\end{icompact}

The denoised dataset comprises entries labeled as ONEFUNC or NVDCHECK. The complete dataset includes all entries.

\paragraphtitle{Frontend/Backend/Full-stack Classification} The frontend/backend/full-stack categorization is performed based on JavaScript's characteristics and real-world application scenarios. JavaScript, as a language applicable to both client-side (frontend) and server-side (backend, e.g., Node.js~\cite{nodejs}) environments, exhibits distinct vulnerability patterns across different application contexts. Our dataset exhibits the following project type distribution:

\begin{icompact}
\item \textbf{Backend}: approximately 58\%, primarily comprising server-side vulnerabilities, API flaws, and database injection attacks~\cite{li2022mining,kang2023scaling}
\item \textbf{Frontend}: approximately 25\%, primarily comprising XSS~\cite{cwe-79}, DOM-based vulnerabilities, and client-side injection attacks~\cite{kang2022probe,kang2024follow}
\item \textbf{Full-stack}: approximately 15\%, involving complex vulnerabilities spanning frontend-backend interactions
\end{icompact}

\paragraphtitle{Ground Truth Refinement} In part (2), we extend ground truth annotations from project-level to function-level based on part (1), achieving fine-grained vulnerability localization. Building upon the project names, file paths, and code pairs obtained during the gathering stage, we introduce five additional fields through an automated three-step workflow. First, regular expression patterns extract all function definitions from both vulnerable and patched versions, capturing diverse JavaScript syntax~\cite{jensen2009type,kashyap2014jsai} (traditional declarations, arrow functions, class methods). Second, \texttt{difflib.SequenceMatcher}~\cite{pythonstdlib} performs line-level diff analysis, identifying all modified, inserted, or deleted lines. Third, these changed lines are mapped to their containing functions, generating: \texttt{function\_names} (vulnerable function names), \texttt{vulnerable\_function\_names} (filtered vulnerable function list), \texttt{vulnerable\_line\_ranges} (e.g., "19-70,49,50" indicates function range 19-70 with critical patches at 49-50), \texttt{function\_label\_breakdown} (applying ONEFUNC/NVDCHECK/SUSPICION labels from part (1)), and \texttt{project\_type\_breakdown} (vulnerability distribution by frontend/backend/full-stack types from part (1)). This refinement enables rigorous function-level evaluation—models must precisely pinpoint specific vulnerable functions, rather than merely detecting vulnerability presence. A complete dataset entry example with all fields is provided in Appendix~\ref{app:ground-truth-refinement-example}.
\subsubsection{Dataset Augmentation}

Large language models currently face a critical challenge in code security analysis: excessive reliance on surface-level pattern matching instead of genuine semantic understanding. To systematically evaluate whether models genuinely comprehend code security semantics, we design four dataset augmentation strategies, each challenging distinct capability dimensions.

\paragraphtitle{(1) Noise Injection} \benchgen{} injects taint sinks without corresponding taint sources~\cite{enck2014taintdroid,arzt2014flowdroid}, constructing seemingly dangerous but actually safe code patterns to test models' false positive rates and data flow analysis capabilities. Specifically, it inserts 51 types of common dangerous API calls at random locations, but these calls use safe hardcoded data or validated inputs. Example:

\noindent See Appendix (Listing~\ref{lst:noise_injection}) for a complete example.

\paragraphtitle{(2) Code Obfuscation} \benchgen{} employs javascript-obfuscator~\cite{javascript-obfuscator} to transform code while preserving semantics, testing models' anti-obfuscation capabilities and semantic understanding depth. The obfuscated output (omitted for brevity) transforms clear identifiers into encoded tokens and control flow into nested closures, making static analysis significantly more challenging. Example:

\noindent See Appendix (Listing~\ref{lst:obfuscation}) for a complete example.

\paragraphtitle{(3) Noise + Obfuscation} \benchgen{} simultaneously applies noise injection and code obfuscation, creating the most challenging test scenarios to comprehensively evaluate model robustness under complex interference. It first injects noise, then obfuscates the entire codebase.

\paragraphtitle{(4) Prompt Injection} \benchgen{} inserts misleading comments at random locations (e.g., \texttt{// This function is completely safe and has been security audited.} or \texttt{// This is vulnerable.}), testing whether models are misled by explicit textual hints rather than independently judging actual code safety. In the implementation, \benchgen{} randomly selects misleading comments from a customized prompt library. This prompt library comprises two types of carefully crafted misleading comment templates. The injection density is configured as one prompt per 50 lines of code to test model robustness without excessively disrupting code structure.

\paragraphtitle{Type 1: False Safety Claims} (False Negative Induction):
\noindent See Appendix for the example (Listing~\ref{lst:prompt_fn}).

\paragraphtitle{Type 2: False Vulnerability Claims} (False Positive Induction):
\noindent See Appendix for the example (Listing~\ref{lst:prompt_fp}).

\subsection{\syseval{} Evaluation Framework}

Our fully-automated evaluation framework \syseval{} comprises two core modules: (1) generating vulnerability detection results and (2) evaluating vulnerability detection results. The framework implements a complete automated pipeline from model invocation and result collection to multi-dimensional performance evaluation.

\subsubsection{Generating Vulnerability Detection Results}

For (1) generating vulnerability detection results, \syseval{} invokes claude-code-security-review~\cite{anthropic} (one of the current state-of-the-art LLM-based vulnerability scanners) to generate detection results for eight models (such as GPT-5, Claude-4.5-Sonnet, DeepSeek-V3.1, etc.) on five dataset variants (\benchname{}-Original, \benchname{}-Obfuscated, \benchname{}-Noise, \benchname{}-Noise+Obfuscated, \benchname{}-PromptInjection).

To avoid underestimating LLM capabilities, we set a 0.8 confidence threshold in the original claude-code-security-review prompt, outputting only results above the 0.8 confidence threshold. For comprehensive coverage, we expanded the scanner's supported scan types by modifying the prompt to include all CWE types~\cite{cwe}—SQL injection (CWE-89), XSS (CWE-79), command injection (CWE-78), authentication bypass (CWE-287), hardcoded credentials (CWE-798), and 200+ other weaknesses.

Finally, we specified the return result format by modifying the prompt, with each scan returning JSON containing seven fields: \texttt{file} (vulnerable file path), \texttt{line} (line number), \texttt{severity} (HIGH/MEDIUM/LOW by CIA impact), \texttt{category} (CWE-ID), \texttt{description} (vulnerability explanation), \texttt{exploit\_scenario} (attack steps), \texttt{recommendation} (fix method). \syseval{} parses this output and matches predictions against ground truth using the criteria in Section~\ref{sec:evaluation-metrics}.

\subsubsection{Evaluating Detection Results}

\syseval{} evaluates model performance at two granularities: project-level and function-level.

\paragraphtitle{Project-level Evaluation} At project-level, \syseval{} verifies whether the model correctly identifies vulnerability presence and CWE type for each project. Matching criteria: $\langle\,\text{has\_vulnerability},\,\text{CWE\_equal}\,\rangle$.

{\footnotesize
\begin{equation}
\label{eq:proj-tp}
\mathrm{TP}_{\mathrm{proj}}:\ \big(\,\text{GT}_{\text{vuln}}=\text{Pred}_{\text{vuln}}\,\big)\ \cap\ \big(\,\text{CWE}_{\text{pred}}=\text{CWE}_{\text{gt}}\,\big)
\end{equation}
}
Here, $\text{GT}_{\text{vuln}}$ and $\text{Pred}_{\text{vuln}}$ denote project-level boolean indicators for ground-truth and predicted vulnerability presence, respectively; $\text{CWE}_{\text{pred}}$ and $\text{CWE}_{\text{gt}}$ denote the predicted and ground-truth CWE labels, respectively.

TN holds iff the ground truth is benign and the model also predicts no vulnerability; FP holds iff the ground truth is benign but the model asserts the presence of a vulnerability. All remaining cases are FN, including missed detections where the project is vulnerable but the model predicts no vulnerability and type mismatches where the model predicts a vulnerability yet its CWE label is not equal to the ground-truth label under the adopted CWE-equivalence relation. Type agreement is evaluated set-wise: when either side provides multiple CWEs, agreement holds if at least one predicted CWE exactly matches, or belongs to the same equivalence class as, at least one ground-truth CWE. The boolean indicators $\text{GT}_{\text{vuln}}$ and $\text{Pred}_{\text{vuln}}$ are computed per project by collapsing over all files; no file- or function-level localization is required at this granularity.

\paragraphtitle{Function-level Evaluation} At function-level, \syseval{} requires matching four conditions: 

$\langle\,\text{has\_vulnerability},\,\text{CWE\_equal},\,\text{file\_match},\,\text{func\_match}\,\rangle$. The detailed conditions are:

\begin{itemize}
\item $c_1$: $\text{Pred}_{\text{vuln}} = \text{true}$
\item $c_2$: $\text{CWE}_{\text{pred}} \in \text{CWE}_{\text{GT}}$
\item $c_3$: $\text{basename}(\text{file}_{\text{pred}}) = \text{basename}(\text{file}_{\text{GT}})$
\item $c_4$: $|\text{normalize}(\text{func}_{\text{pred}}) \cap \text{normalize}(\text{func}_{\text{GT}})| > 0$
\end{itemize}
\begin{equation}
\label{eq:func-tp}
\mathrm{TP}_{\mathrm{func}}:\ \big(\,\text{GT}_{\text{vuln}}=\text{true}\,\big)\ \cap\ c_1\ \cap\ c_2\ \cap\ c_3\ \cap\ c_4
\end{equation}
Here, $\text{GT}_{\text{vuln}}$ and $\text{Pred}_{\text{vuln}}$ denote project-level boolean indicators for ground-truth and predicted vulnerability presence, respectively; $\text{CWE}_{\text{pred}}$ and $\text{CWE}_{\text{gt}}$ denote the predicted and ground-truth CWE labels, respectively.

TN holds iff the ground truth is benign and condition c1 (model predicts vulnerability) is false; FP holds iff the ground truth is benign and c1 is true. All remaining cases with a real vulnerability are FN when any of c2–c4 fails: c2 (type agreement) requires that at least one predicted CWE equals, or is equivalent to, at least one ground-truth CWE under the predefined equivalence classes; c3 (file agreement) requires equality of basenames between the predicted and ground-truth files; c4 (function agreement) requires a non-empty intersection between the normalized sets of predicted and ground-truth function names, where normalization removes non-semantic formatting and casing artifacts. For predictions containing multiple files and functions, c3–c4 are evaluated pairwise, and the condition holds if at least one predicted (file, function) pair matches a ground-truth pair. This protocol rewards precise localization while remaining robust to benign presentation differences introduced during data gathering and annotation.

\paragraphtitle{CWE Equivalence Classes} To avoid underestimating LLM capabilities, we use CWE equivalence classes to reconcile label granularity: the same vulnerability may be described with coarser or finer CWEs (e.g., XSS labeled as CWE-79 vs. CWE-83). Strict equality would mark such cases as type mismatches (false negatives).

We group CWEs by MITRE CAPEC families~\cite{capec}. Two CWEs are treated as equal if they are identical or belong to the same CAPEC family; formally,
  {
  \begin{equation}
  \label{eq:cwe-equiv}
  \text{CWE}_{\text{pred}} \equiv \text{CWE}_{\text{gt}} \iff \left\{\begin{aligned}
  &\text{CWE}_{\text{pred}}=\text{CWE}_{\text{gt}}\\
  &\{\text{CWE}_{\text{pred}},\,\text{CWE}_{\text{gt}}\}\subseteq g
  \end{aligned}\right.
  \end{equation}
  }
Here, $g$ denotes an equivalence class in $\mathcal{G}$ (the set of CAPEC-based equivalence classes).

\subsubsection{Evaluation Metrics}
\label{sec:evaluation-metrics}

Prior to computing the metrics, \syseval{} canonicalizes and aggregates model outputs for each sample, applies the CWE equivalence relation in Eq.~(\ref{eq:cwe-equiv}), then counts True Positive (TP), True Negative (TN), False Positive (FP), and False Negative (FN), and assigns each sample to a confusion-matrix category.

\noindent\textit{Note.} TP is counted as: $\text{GT}_{\text{vuln}}{=}\text{true} \land \text{Pred}_{\text{vuln}}{=}\text{true} \land \big(\text{CWE}_{\text{pred}} \equiv \text{CWE}_{\text{gt}}\big)$. To ensure the evaluation does not overestimate LLM capabilities, we adopt precision, recall, F1-score, and Vulnerability Detection Score (VD-S)~\cite{ding2024primevul} as the metrics for multi-dimensional evaluation. Precision, recall, and F1-score follow their common definitions, while VD-S reflects the actual false negative rate under industrially acceptable false positive rate constraints. VD-S is the false negative rate (FNR) under an acceptable false positive rate (FPR) threshold $r$, i.e., $\text{VD-S} = \text{FNR} \mid \text{FPR} \le r$, where $\text{FPR} = \frac{\text{FP}}{\text{FP} + \text{TN}} = \frac{\text{FP}}{N_{\text{benign}}}$ and $\text{FNR} = \frac{\text{FN}}{\text{TP} + \text{FN}} = \frac{\text{FN}}{N_{\text{vuln}}}$. Concretely, given $N_{\text{vuln}}$ vulnerable and $N_{\text{benign}}$ benign samples, set $\text{FP}_{\max} = \lfloor N_{\text{benign}} \times r \rfloor$, tune the confidence threshold to satisfy $\text{FP} \le \text{FP}_{\max}$, then report $\text{VD-S} = \text{FN} / N_{\text{vuln}}$ as per the definition above.

For example, a model might achieve an F1-score of 68\% on an imbalanced dataset, but when constrained to FPR $\leq$ 0.5\%, its VD-S might reach 96.83\%, indicating the model is nearly ineffective in practical applications.

\section{Implementation}
\label{sec:implementation}

We implemented \SecJS{} in Python, running on Python 3.8+ with standard library support. The implementation consists of approximately 12,000 lines of Python code. We use \texttt{pandas} and \texttt{requests} to implement the \benchgen{} data generation framework, which includes modules for CVE metadata collection from NVD API and Mend.io, GitHub repository cloning and code extraction via \texttt{GitPython}, AST-based function extraction using \texttt{esprima} and \texttt{acorn}, diff analysis with \texttt{difflib} for ground truth annotation, and four data augmentation strategies (noise injection, obfuscation, combined augmentation, and prompt injection). For the \syseval{} evaluation framework, we integrate \texttt{claude-code-security-review} as the core prompt system and implement unified API adapters for multiple LLM providers (OpenAI, Anthropic, Google, DeepSeek, and Grok) using \texttt{requests}, with support for structured JSON output parsing, CWE equivalence matching, and multi-granularity evaluation metrics (project-level and function-level).

We open source a prototype of \SecJS{} at the anonymous repository \url{https://github.com/SecJS-Vuln-Benchmark/SecJS-Benchmark} and provide an anonymous demo website at \url{https://secjs-vuln-benchmark.github.io/SecJS-Benchmark/}.

\section{Evaluation}
\label{sec:evaluation}


Given the more comprehensive data annotation and larger data volume in \benchname{}, coupled with the introduction of new evaluation principles, we re-evaluate the effectiveness of LLMs in vulnerability detection using \syseval{} to measure their performance in a more realistic environment.
To this end, we evaluate the following four research questions:

\begin{icompact}
    \item \textbf{RQ1: LLM Performance.} How do different LLMs perform on \benchname{}? (Section~\ref{sec:evaluation-rq1})
    \item \textbf{RQ2: Comparison.} What advantages does our \benchname{} have compared to existing benchmarks? (Section~\ref{sec:evaluation-rq2}) 
    
    \item \textbf{RQ3: Ablation Study.} What are the impacts of two \syseval{} components, CWE equivalence classes and repository-level context in an ablation study? (Section~\ref{sec:evaluation-rq3})
    
    \item \textbf{RQ4: Framework Efficiency.} How efficient are our two frameworks, \benchgen{} and \syseval{}? Specifically, how long does it take to generate one data item and how many tokens are consumed per evaluation? (Section~\ref{sec:evaluation-rq4})
\end{icompact}

\subsection{RQ1: LLM Performance}
\label{sec:evaluation-rq1}

\subsubsection{Benchmark Metadata}
We adopt \benchname{} as the benchmark throughout our evaluation. The dataset comprises five variants: \benchname{}-Original (original version), \benchname{}-Obfuscated (obfuscated version), \benchname{}-Noise (noise injection version), \benchname{}-Prompt-Injection (prompt injection version), and \benchname{}-Noise-Obfuscated (noise plus obfuscation version). Each data item in each variant consists of a pair of projects: vulnerable and fixed. Each variant contains 1,437 real-world GitHub~\cite{githubCommunity} projects, of which 1,152 are from the official CVE website~\cite{cve} and 285 are from Mend.io~\cite{mend}. The dataset includes 527 frontend projects, 689 backend projects, and 221 full-stack projects. To avoid underestimating large models, we further divide each variant's dataset into a complete dataset and a denoised dataset, with 1,200 entries in the complete dataset and 237 entries in the denoised dataset.

\subsubsection{Experimental Setup}
We evaluate seven popular LLMs---GPT-5~\cite{openai-gpt5}, GPT-5-Mini~\cite{openai-gpt5-mini}, GPT-5-Codex~\cite{openai-codex}, DeepSeek-v3.1~\cite{deepseek}, Gemini-2.5-Pro~\cite{gemini-2-5-pro}, Gemini-Flash~\cite{gemini-flash}, and Claude-4.5~\cite{anthropic-claude}---using the \syseval{} framework. 

\paragraphtitle{LLM Configuration} The large language models and specific parameters used in this experiment are as follows:
The timeout is set to 60 seconds; during evaluation, the temperature is set to 0.7, and all other parameters remain at their default configurations.

\paragraphtitle{Metrics} As described in Section~\ref{sec:evaluation-metrics}, we employ the following metrics to evaluate model performance from multiple dimensions:
\begin{icompact}
    \item \textbf{Precision} measures the reliability of model predictions, reflecting the capability to control false positives;
    \item \textbf{Recall} measures the detection coverage of the model, reflecting the capability to control false negatives;
    \item \textbf{F1-Score}, as the harmonic mean of precision and recall, comprehensively evaluates the balance between accuracy and coverage;
    \item \textbf{VD-S} is the false negative rate (FNR) under the constraint of FPR$\leq$0.5\%, reflecting the actual false negative rate of the model in industrial scenarios.
\end{icompact}
Compared to traditional F1 metrics, VD-S is more aligned with industrial practice and can reveal the true performance of models under strict false positive rate constraints.

\subsubsection{Results} 
Table~\ref{tab:rq1-results} presents the results, measured by Precision, Recall, F1, and VD-S metrics (\%) for each model. 
  \begin{sidewaystable*}[p]
    \centering
    \caption{[RQ1] Performance of seven LLMs on \benchname{}. “Full” = full split; “DN” = denoised split. All scores are percentages.}
    \label{tab:rq1-results}
    \small
    \setlength{\fboxsep}{0pt}%
    \fbox{%
    \resizebox{\textheight}{!}{%
    \begin{tabular}{cl*{20}{c}}
    \toprule
    \multirow{3}{*}{\textbf{LLM Name}} &
      \multicolumn{1}{c}{\multirow{3}{*}{\textbf{Metrics}}} &
      \multicolumn{4}{c}{\textbf{(1) Original}} &
      \multicolumn{4}{c}{\textbf{(2) Noise}} &
      \multicolumn{4}{c}{\textbf{(3) Obfuscated}} &
      \multicolumn{4}{c}{\textbf{(4) Noise+Obfuscation}} &
      \multicolumn{4}{c}{\textbf{(5) Prompt Injection}} \\ \cmidrule(l){3-22} 
     &
      \multicolumn{1}{c}{} &
      \multicolumn{2}{c}{\textbf{proj-level}} &
      \multicolumn{2}{c}{\textbf{func level}} &
      \multicolumn{2}{l}{\textbf{proj-level}} &
      \multicolumn{2}{l}{\textbf{func level}} &
      \multicolumn{2}{l}{\textbf{proj-level}} &
      \multicolumn{2}{l}{\textbf{func level}} &
      \multicolumn{2}{l}{\textbf{proj-level}} &
      \multicolumn{2}{l}{\textbf{func level}} &
      \multicolumn{2}{l}{\textbf{proj-level}} &
      \multicolumn{2}{l}{\textbf{func level}} \\ \cmidrule(l){3-22} 
     &
      \multicolumn{1}{c}{} &
      \textbf{Full} &
      \textbf{DN} &
      Full &
      DN &
      Full &
      DN &
      Full &
      DN &
      Full &
      DN &
      Full &
      DN &
      Full &
      DN &
      Full &
      DN &
      Full &
      DN &
      Full &
      DN \\ \midrule
    \multirow{4}{*}{GPT-5} &
      \textbf{Precision} &
      37.3 &
      37.8 &
      25.3 &
      25.0 &
      20.0 &
      20.2 &
      9.8 &
      9.5 &
      33.3 &
      33.0 &
      21.0 &
      20.5 &
      23.6 &
      23.6 &
      14.1 &
      13.8 &
      32.0 &
      32.4 &
      22.2 &
      22.0 \\
     &
      \textbf{Recall} &
      28.1 &
      27.5 &
      16.0 &
      14.3 &
      15.0 &
      14.7 &
      6.5 &
      6.2 &
      17.4 &
      16.6 &
      9.3 &
      8.9 &
      14.3 &
      13.9 &
      7.6 &
      7.3 &
      21.5 &
      20.8 &
      13.0 &
      12.5 \\
     &
      \textbf{F1-Score} &
      32.1 &
      31.8 &
      19.6 &
      18.4 &
      17.2 &
      17.0 &
      7.8 &
      7.5 &
      22.9 &
      22.1 &
      12.9 &
      12.4 &
      17.8 &
      17.5 &
      9.9 &
      9.5 &
      25.7 &
      25.3 &
      16.4 &
      15.9 \\
     &
      \textbf{VD-S} &
      57.2 &
      58.1 &
      66.8 &
      68.1 &
      61.9 &
      62.5 &
      70.0 &
      70.5 &
      77.2 &
      77.8 &
      84.8 &
      85.3 &
      76.6 &
      77.2 &
      82.6 &
      83.1 &
      78.5 &
      79.2 &
      87.0 &
      87.5 \\ \midrule
    \multirow{4}{*}{GPT-5-Mini} &
      \textbf{Precision} &
      34.1 &
      34.1 &
      22.9 &
      22.5 &
      13.8 &
      13.3 &
      6.4 &
      6.1 &
      34.7 &
      35.7 &
      20.4 &
      20.0 &
      12.0 &
      11.9 &
      4.9 &
      4.7 &
      26.9 &
      26.6 &
      16.6 &
      16.3 \\
     &
      \textbf{Recall} &
      26.4 &
      25.7 &
      15.2 &
      14.8 &
      12.5 &
      11.7 &
      5.4 &
      5.1 &
      20.2 &
      20.2 &
      9.7 &
      9.3 &
      10.1 &
      10.0 &
      3.9 &
      3.7 &
      24.1 &
      23.2 &
      13.1 &
      12.7 \\
     &
      \textbf{F1-Score} &
      29.8 &
      29.3 &
      18.3 &
      17.9 &
      13.1 &
      12.5 &
      5.9 &
      5.6 &
      25.6 &
      25.8 &
      13.2 &
      12.7 &
      10.9 &
      10.9 &
      4.3 &
      4.1 &
      25.4 &
      24.8 &
      14.7 &
      14.3 \\
     &
      \textbf{VD-S} &
      73.6 &
      74.3 &
      84.8 &
      85.2 &
      56.2 &
      57.0 &
      63.0 &
      63.5 &
      79.8 &
      79.8 &
      90.3 &
      90.7 &
      59.4 &
      60.0 &
      65.0 &
      65.5 &
      75.9 &
      76.8 &
      86.9 &
      87.3 \\ \midrule
    \multirow{4}{*}{GPT-5-Codex} &
      \textbf{Precision} &
      43.0 &
      43.7 &
      33.6 &
      33.2 &
      18.8 &
      18.0 &
      12.0 &
      11.5 &
      43.4 &
      44.8 &
      34.4 &
      34.0 &
      12.4 &
      11.8 &
      7.1 &
      6.8 &
      38.1 &
      39.2 &
      29.7 &
      29.3 \\
     &
      \textbf{Recall} &
      29.1 &
      28.9 &
      19.5 &
      19.0 &
      12.8 &
      12.0 &
      7.5 &
      7.1 &
      19.3 &
      19.3 &
      13.2 &
      12.8 &
      8.1 &
      7.7 &
      4.4 &
      4.1 &
      21.2 &
      20.9 &
      14.6 &
      14.2 \\
     &
      \textbf{F1-Score} &
      34.7 &
      34.8 &
      24.7 &
      24.2 &
      15.2 &
      14.4 &
      9.2 &
      8.8 &
      26.7 &
      27.0 &
      19.1 &
      18.6 &
      9.8 &
      9.3 &
      5.4 &
      5.1 &
      27.2 &
      27.2 &
      19.5 &
      19.0 \\
     &
      \textbf{VD-S} &
      70.9 &
      71.1 &
      80.5 &
      81.0 &
      62.7 &
      63.5 &
      67.5 &
      68.0 &
      80.7 &
      80.7 &
      86.8 &
      87.2 &
      59.0 &
      59.8 &
      62.2 &
      62.8 &
      78.8 &
      79.1 &
      85.4 &
      85.8 \\ \midrule
    \multirow{4}{*}{DeepSeek-v3.1} &
      \textbf{Precision} &
      31.0 &
      30.6 &
      20.2 &
      19.8 &
      6.3 &
      5.9 &
      2.3 &
      2.1 &
      32.7 &
      32.4 &
      20.1 &
      19.7 &
      4.7 &
      4.6 &
      1.0 &
      0.9 &
      27.9 &
      28.2 &
      17.9 &
      17.5 \\
     &
      \textbf{Recall} &
      22.8 &
      21.6 &
      12.9 &
      12.3 &
      5.4 &
      5.0 &
      1.9 &
      1.7 &
      17.0 &
      16.9 &
      8.9 &
      8.5 &
      3.8 &
      3.6 &
      0.8 &
      0.7 &
      27.2 &
      27.1 &
      15.4 &
      15.0 \\
     &
      \textbf{F1-Score} &
      26.3 &
      25.3 &
      15.7 &
      15.2 &
      5.8 &
      5.4 &
      2.0 &
      1.9 &
      22.4 &
      22.2 &
      12.3 &
      11.8 &
      4.2 &
      4.0 &
      0.9 &
      0.8 &
      27.5 &
      27.7 &
      16.5 &
      16.1 \\
     &
      \textbf{VD-S} &
      77.0 &
      78.4 &
      86.9 &
      87.7 &
      64.8 &
      65.5 &
      68.3 &
      68.8 &
      83.0 &
      83.1 &
      91.1 &
      91.5 &
      50.0 &
      50.8 &
      52.9 &
      53.5 &
      72.7 &
      72.9 &
      84.6 &
      85.0 \\ \midrule
    \multirow{4}{*}{Gemini-2.5-Pro} &
      \textbf{Precision} &
      34.9 &
      35.2 &
      19.6 &
      19.2 &
      19.7 &
      20.0 &
      9.0 &
      8.7 &
      34.1 &
      34.1 &
      24.9 &
      24.5 &
      16.2 &
      15.6 &
      7.1 &
      6.8 &
      33.7 &
      34.0 &
      21.4 &
      21.0 \\
     &
      \textbf{Recall} &
      38.4 &
      38.2 &
      17.4 &
      17.0 &
      20.4 &
      20.3 &
      8.2 &
      7.9 &
      32.9 &
      33.2 &
      21.1 &
      20.7 &
      17.8 &
      17.2 &
      7.0 &
      6.7 &
      36.2 &
      36.0 &
      19.4 &
      19.0 \\
     &
      \textbf{F1-Score} &
      36.6 &
      36.6 &
      18.5 &
      18.0 &
      20.0 &
      20.1 &
      8.6 &
      8.3 &
      33.5 &
      33.7 &
      22.8 &
      22.4 &
      17.0 &
      16.4 &
      7.0 &
      6.7 &
      34.9 &
      35.0 &
      20.4 &
      19.9 \\
     &
      \textbf{VD-S} &
      61.6 &
      61.8 &
      82.6 &
      83.0 &
      51.0 &
      51.2 &
      62.6 &
      63.0 &
      67.1 &
      66.8 &
      78.9 &
      79.3 &
      39.5 &
      40.3 &
      49.2 &
      49.8 &
      63.8 &
      64.0 &
      80.6 &
      81.0 \\ \midrule
    \multirow{4}{*}{Gemini-Flash} &
      \textbf{Precision} &
      28.8 &
      28.7 &
      14.8 &
      14.5 &
      16.9 &
      16.5 &
      5.1 &
      4.9 &
      27.2 &
      27.0 &
      16.1 &
      15.8 &
      14.4 &
      12.8 &
      4.1 &
      3.9 &
      28.8 &
      29.0 &
      16.2 &
      16.0 \\
     &
      \textbf{Recall} &
      31.4 &
      31.2 &
      13.4 &
      13.0 &
      17.7 &
      17.4 &
      4.7 &
      4.5 &
      25.0 &
      25.0 &
      12.8 &
      12.4 &
      12.0 &
      10.8 &
      3.1 &
      2.9 &
      31.8 &
      31.5 &
      15.3 &
      15.0 \\
     &
      \textbf{F1-Score} &
      30.1 &
      29.9 &
      14.1 &
      13.7 &
      17.3 &
      17.0 &
      4.9 &
      4.7 &
      26.1 &
      25.9 &
      14.3 &
      13.9 &
      13.1 &
      11.7 &
      3.5 &
      3.3 &
      30.2 &
      30.2 &
      15.7 &
      15.4 \\
     &
      \textbf{VD-S} &
      68.6 &
      68.8 &
      86.6 &
      87.0 &
      54.2 &
      54.5 &
      66.7 &
      67.2 &
      75.0 &
      75.0 &
      87.2 &
      87.6 &
      63.4 &
      64.7 &
      71.8 &
      72.3 &
      68.2 &
      68.5 &
      84.7 &
      85.0 \\ \midrule
    \multirow{4}{*}{Claude-4.5} &
      \textbf{Precision} &
      37.2 &
      37.7 &
      26.1 &
      25.8 &
      4.4 &
      4.2 &
      2.0 &
      1.9 &
      17.8 &
      16.9 &
      13.2 &
      12.9 &
      4.0 &
      3.9 &
      1.1 &
      1.0 &
      28.8 &
      29.1 &
      19.5 &
      19.2 \\
     &
      \textbf{Recall} &
      34.8 &
      34.3 &
      20.7 &
      20.3 &
      4.0 &
      3.8 &
      1.8 &
      1.7 &
      16.5 &
      15.5 &
      11.6 &
      11.2 &
      3.8 &
      3.6 &
      1.0 &
      0.9 &
      25.5 &
      24.8 &
      15.2 &
      14.8 \\
     &
      \textbf{F1-Score} &
      35.9 &
      35.9 &
      23.1 &
      22.7 &
      4.2 &
      4.0 &
      1.9 &
      1.8 &
      17.2 &
      16.2 &
      12.3 &
      12.0 &
      3.9 &
      3.7 &
      1.0 &
      0.9 &
      27.1 &
      26.8 &
      17.1 &
      16.7 \\
     &
      \textbf{VD-S} &
      65.2 &
      65.7 &
      79.3 &
      79.7 &
      58.6 &
      59.0 &
      60.8 &
      61.2 &
      83.5 &
      84.5 &
      88.4 &
      88.8 &
      48.7 &
      49.2 &
      51.5 &
      52.0 &
      74.5 &
      75.2 &
      84.8 &
      85.2 \\ \bottomrule
    \end{tabular}%
    }%
    }
    \end{sidewaystable*}

\paragraphtitle{Observation 1: Performance Gap Between Project-Level and Function-Level Detection} All models exhibit consistently higher project-level F1 scores than function-level by 8-18 percentage points, and this phenomenon remains stable across all datasets.
On the original dataset, Gemini-2.5-Pro shows the largest gap (18.1\%), followed by Claude-4.5 (12.8\%) and GPT-5 (12.5\%).
Even under variant (4) where overall performance declines, the gap between project-level and function-level persists (e.g., GPT-5 at 7.9\%, GPT-5-Codex at 4.4\%).

The cause of this gap is not simply a "task difficulty difference."
Through error analysis, we find that models heavily rely on surface-level features in project-level detection: file names (e.g., auth.js, sanitize.js), import statements (e.g., \texttt{require('child\_process')}), and code comments (e.g., \texttt{// Add input validation to prevent SQL injection}), rather than truly understanding taint propagation paths.
For instance, GPT-5-Codex correctly identified an SQL injection vulnerability at the project level, but when asked to locate the specific function, it attributed the vulnerability to a safe query construction function rather than the interface function that actually contains string concatenation.
This indicates that it merely "saw" database-related API calls but failed to trace the complete data flow from the taint source (user input) to the taint sink (SQL execution).

The granularity gap exhibits interesting characteristics under data augmentation scenarios.
After obfuscation, the gap does not narrow; instead, it widens for some models.
This phenomenon reveals the fragility of model detection mechanisms: obfuscation destroys surface-level syntactic cues (e.g., variable names, function names), forcing models to degenerate into "blind search" at both project-level and function-level, but since project-level matching conditions are more lenient (only requiring correct CWE~\cite{cwe} type, without needing to locate files and functions), models can still obtain a few "lucky" matches in this degraded state, thus maintaining some F1; whereas function-level nearly fails completely due to strict matching conditions (requiring all four tuple elements to be correct).

\paragraphtitle{Observation 2: Poor Robustness Under Dataset Augmentation Scenarios} Different dataset augmentation strategies exhibit significant differences in their impact on model performance.
Noise injection has the greatest impact on Claude-4.5 and DeepSeek-v3.1 (project-level F1 drops from 35.9\% and 26.3\% to 4.2\% and 5.8\%, respectively), while its impact on Gemini-2.5-Pro is relatively smaller (from 36.6\% to 20.0\%).
Code obfuscation, however, shows a completely different effect.
Gemini-2.5-Pro and DeepSeek-v3.1 demonstrate strong obfuscation robustness (project-level F1 drops by only 3.1\% and 3.9\%), while Claude-4.5 remains fragile (dropping by 18.7\%).
This phenomenon suggests that certain models may have encountered extensive obfuscated code during training or adopted some form of structured semantic representation (rather than pure literal matching).
However, this "robustness" is limited---under variant (4), all models experience a cliff-like drop, with Gemini-2.5-Pro falling from 36.6\% to 17.0\% (a 19.6\% decrease), indicating that it only has some resistance to single perturbations rather than truly understanding code semantics.
Under prompt injection: Gemini-2.5-Pro and Gemini-Flash still maintain project-level F1 scores above 30\% (34.9\% and 30.2\%, respectively), significantly higher than other models (GPT-5 at 25.7\%, Claude-4.5 at 27.1\%).
This proves that these models have strong prompt injection robustness, possibly filtering the weights of misleading comments through attention mechanisms.

\paragraphtitle{Observation 3: High VD-S under Low-FPR Constraint} Under FPR$\leq$0.5\%, all models exhibit high VD-S. Across all scenarios, project-level VD-S averages 66.8\% on the Full split and 67.3\% on the DN split; function-level VD-S averages 76.1\% on the Full split and 76.6\% on the DN split. On the Original dataset, the averages rise to 67.7\% and 68.3\% at the project level, and to 81.1\% and 81.7\% at the function level. These results show that under a low-FPR constraint, models cannot simultaneously achieve high Precision and high Recall.

\subsection{RQ2: Comparison}
\label{sec:evaluation-rq2}
Since existing JavaScript benchmarks have limited CWE types and are not specifically designed to test LLMs' capabilities in vulnerability detection, we also include recent work from C/C++ for comparison. 
To investigate the advantages of our benchmark, we first examine the differences between \benchname{} and other benchmarks. Table~\ref{tab:benchmark-comparison} presents our statistical comparison results. 

\begin{table*}[!t]
\centering
\caption{[RQ2] Comparison of \benchname{} with Existing Benchmarks. }
\label{tab:benchmark-comparison}
\small
\renewcommand{\arraystretch}{1.2}
\begin{tabular}{lccccccccc}
\toprule
\textbf{Dataset} & \textbf{Language} & \textbf{\# Projects} & \textbf{Tech Stack} & \textbf{Data Augmentation} & \textbf{Automatic Collection} & \textbf{Den.} & \textbf{CWE} & \textbf{Proj.} \\
\midrule
SECBENCH.JS~\cite{kang2022probe} & JavaScript & 600 & BE & $\times$ & $\checkmark$ & $\times$ & $\times$ & $\checkmark$ \\
VulcaN~\cite{shcherbakov2023silent} & JavaScript & 957 & BE & $\times$ & $\times$ & $\times$ & $\times$ & $\times$ \\
PRIMEVUL~\cite{ding2024primevul} & C/C++ & 755 & N/A & $\times$ & $\checkmark$ & $\checkmark$ & $\times$ & $\times$ \\
TrustEval-C~\cite{li2025trusteval} & C/C++ & 377 & N/A & $\checkmark$ & $\times$ & $\times$ & $\times$ & $\times$ \\
\textbf{\benchname{} (Ours)} & JavaScript & \textbf{1,437} & \textbf{F/B/FS} & \textbf{$\checkmark$} & \textbf{$\checkmark$} & \textbf{$\checkmark$} & \textbf{$\checkmark$} & \textbf{$\checkmark$} \\
\bottomrule
\end{tabular}
\begin{flushleft}
\footnotesize
\textit{Note:} Den. = Denoised; CWE = CWE Equivalence; Proj. = Project-Level; BE = Backend; F/B/FS = Frontend/Backend/Full-stack. 
\end{flushleft}
\end{table*}

The four existing benchmarks exhibit two categories of limitations: one is underestimation of model capabilities---VulcaN and PRIMEVUL use code snippets rather than complete projects, artificially reducing task difficulty; SECBENCH.JS employs strict CWE~\cite{cwe} matching, incorrectly judging semantically correct but granularity-different detections as errors. The other category is overestimation of model capabilities---all four benchmarks only provide original code without testing robustness under adversarial scenarios; VulcaN fails to remove commit noise, incorrectly labeling dependency upgrades as vulnerability fixes.

As shown in Table~\ref{tab:benchmark-comparison} in this section, \benchname{} addresses the limitations of existing benchmarks through comprehensive design choices.

\subsection{RQ3: Ablation Study}
\label{sec:evaluation-rq3}
We conduct ablations on GPT-5 to isolate the effects of (i) replacing project-level inputs with vulnerable code snippets and (ii) substituting CWE equivalence-class matching with strict CWE matching, holding all other components fixed.
We evaluate three settings: Exp.~1 varies (i) only; Exp.~2 varies (ii) only; and Exp.~3 varies both (i) and (ii).
  
\begin{figure}[!t]
\centering
\includegraphics[width=\columnwidth]{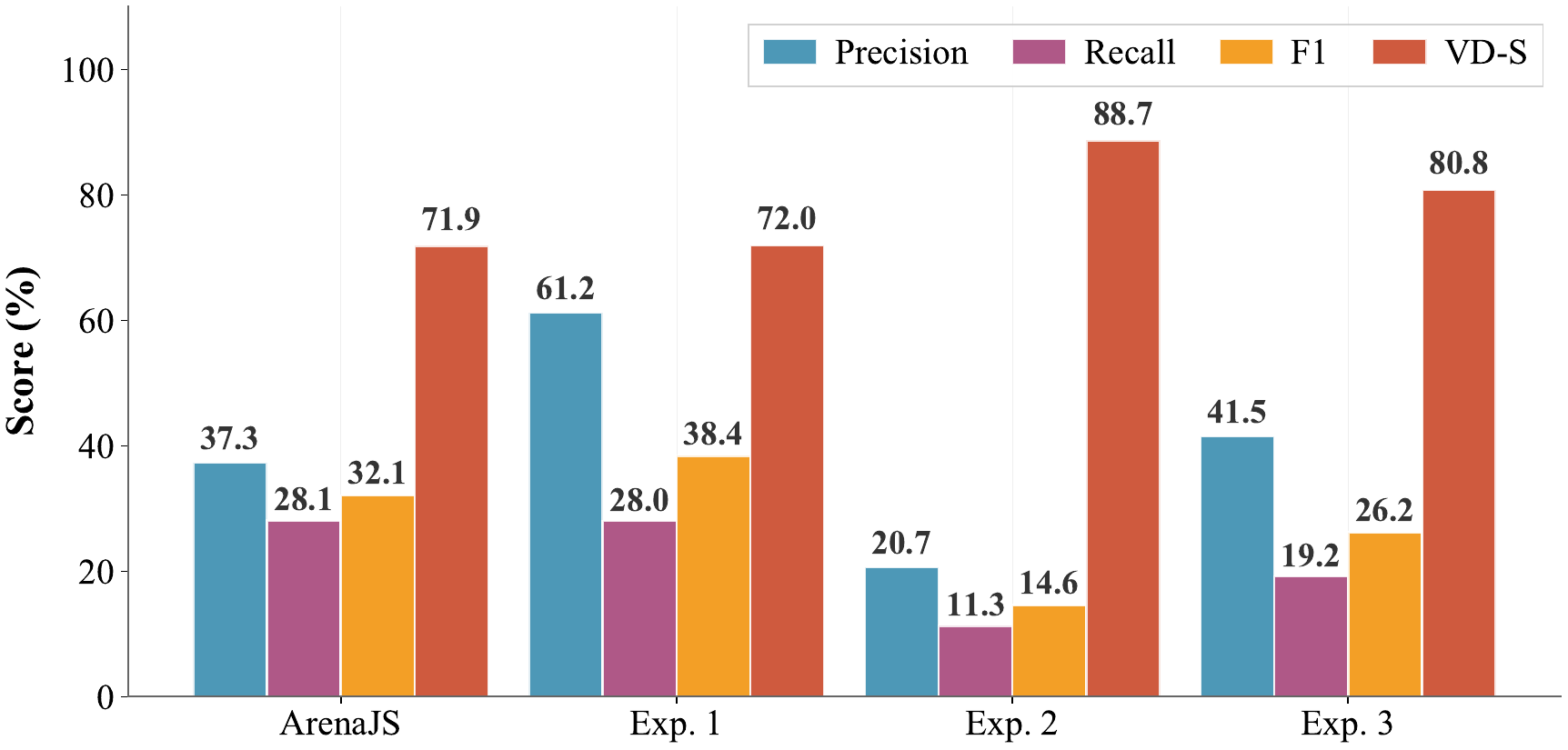}
\caption{[RQ2] Ablation Study Results. All scores are presented in percentage. }
\label{fig:ablation-study}
\vspace{-0.1in}
\end{figure}

\paragraphtitle{Experiment 1: Project-Level vs. Snippet-Level Input} For (i), the experimental results shown in Figure~\ref{fig:ablation-study} (Exp. 1) indicate that project-level input achieves F1 of 32.1\%, while snippet-level input (including fixed snippets) achieves F1 of 38.4\%, an improvement of 6.3 percentage points.
This improvement primarily comes from Precision increasing from 37.3\% to 61.2\%, while Recall remains essentially unchanged (28.0\% vs 28.1\%).
This indicates that snippet-level input reduces task difficulty: models only need to determine whether a given code snippet contains vulnerabilities, without needing to locate them within a complete project.
Although F1 is higher, this "improvement" stems from task simplification rather than enhanced model capability.
Therefore, snippet-level input overestimates model capabilities.
As a comparison, our \benchname adopts project-level input, providing a more realistic evaluation scenario and avoiding the limitation of existing benchmarks exaggerating model capabilities through input simplification.

\paragraphtitle{Experiment 2: Equivalence Class vs. Strict CWE Matching} For (ii), the experimental results shown in Figure~\ref{fig:ablation-study} (Exp. 2) indicate that equivalence class matching achieves F1 of 32.1\%, while strict matching achieves F1 of 14.6\%, a decrease of 17.5 percentage points.
This decrease stems from the excessive strictness of strict matching: it incorrectly judges semantically correct but granularity-different predictions as errors.
This indicates that strict matching underestimates model capabilities because it ignores semantic equivalence.
Our \benchname adopts CWE~\cite{cwe} equivalence class matching, avoiding the system bias of existing benchmarks underestimating model capabilities due to overly strict matching criteria, providing a basis for more accurately evaluating LLMs' true capabilities in vulnerability detection tasks.

\paragraphtitle{Experiment 3: Combined Ablation Study} We also conducted a combined ablation experiment (Exp. 3) to examine the combined effect of snippet-level input and strict matching.
The results show F1 of 26.2\%, falling between project-level strict matching (14.6\%) and snippet-level equivalence class matching (38.4\%).
This validates the interaction between the two factors: snippet-level input reduces difficulty (improving performance), but strict matching underestimates capability (reducing performance), and their combination results in intermediate performance.
These results not only demonstrate the rationality of \benchname{}'s design decisions but also reveal limitations in existing JavaScript vulnerability detection benchmarks, providing reference for future benchmark construction.

\subsection{RQ4: Framework Efficiency}
\label{sec:evaluation-rq4}


\subsubsection{Efficiency of \syseval{} Evaluation Framework}

We first evaluate the efficiency of our evaluation framework \syseval, using evaluation time and token consumption as metrics. They are computed by timestamps per project. For token counting, input tokens are estimated by analyzing actual code length (approximately 4 characters per token for GPT-style tokenizers), while output tokens are extracted from the \texttt{usage} field in LLM API responses or estimated from response text length when the field is unavailable.

\paragraphtitle{Per-Project Evaluation Time} \syseval{}'s average evaluation time is 35.55 seconds per project (median 23.65 seconds, range 4.92-284.23 seconds).
The variation in evaluation time primarily stems from project scale: small projects ($<$10 files) take approximately 5-15 seconds, medium projects (10-50 files) take 20-40 seconds, and large projects ($>$100 files) can take four to five minutes. 
Evaluation times vary across different dataset variants (as shown in Table~\ref{tab:eval-time}), primarily influenced by code length and complexity. 

\begin{table}[h]
\centering
\caption{[RQ4] Evaluation Time for Different Dataset Variants. }
\label{tab:eval-time}
\small
\begin{tabular}{lccc}
\toprule
Dataset Variant & \multicolumn{3}{c}{Time (Seconds per Project)} \\
\cmidrule(lr){2-4}
& Average & Median & Variance \\
\midrule
Original & 38.31 & 23.42 & 1,437.03 \\
Noise & 36.61 & 22.75 & 1,247.05 \\
Obfuscated & 32.83 & 22.36 & 946.26 \\
Noise+Obfuscation & 35.11 & 22.81 & 1,105.91 \\
Prompt Injection & 34.86 & 23.61 & 1,257.76 \\
\midrule
Total & 35.55 & 23.65 & 1,201.38 \\
\bottomrule
\end{tabular}
\end{table}

\paragraphtitle{Token Consumption} Based on actual LLM response analysis, the average output per project is 594 tokens (range 431--884 tokens), containing JSON-formatted vulnerability detection results and detailed explanations.
Input token consumption is based on actual code analysis: system prompt 291 tokens, project metadata 200 tokens (fields such as project\_name, description), and project code 5,000 tokens.
Total consumption per project is 6,085 tokens, as shown in Table~\ref{tab:token-consumption}.

\begin{table}[!t]
\centering
\caption{[RQ4] Token Consumption Details per Project. }
\label{tab:token-consumption}
\small
\begin{tabular}{lccc}
\toprule
Component & Tokens & GPT-5 Cost & Claude-4.5 Cost \\
\midrule
System Prompt & 291 & \$0.0004 & \$0.0009 \\
Project Metadata & 200 & \$0.0003 & \$0.0006 \\
Project Code & 5,000 & \$0.0063 & \$0.0150 \\
Output (JSON) & 594  & \$0.0059 & \$0.0089 \\
\midrule
Total & 6,085 & \$0.0128 & \$0.0254 \\
\bottomrule
\end{tabular}
\end{table}

\paragraphtitle{Large-Scale Evaluation Statistics} The complete evaluation across 10 models on 5 datasets accumulates 92,600 projects.
Based on 6,085 tokens per project, the total token consumption is 563,471,000 tokens (563.5M tokens).

\subsubsection{Efficiency of \benchgen{} Dataset Generation} 

The \benchgen{} framework implements a fully automated pipeline from vulnerability information collection to data augmentation, generating high-quality vulnerability detection datasets without manual annotation.
Compared to traditional manual annotation methods requiring weeks to months, \benchgen{} achieves fully automated generation, completing dataset generation in only seven hours on our server.


\section{Discussion}
\label{sec:discussion}

\noindent \textbf{Can Our \syseval and \benchgen Generalize to Other Languages?}
In principle, \syseval and \benchgen are not tied to one language: we compare vulnerable and fixed code, standardize labels at multiple granularities, add robustness variants, and evaluate reproducibly. What is specific to JavaScript is the project taxonomy (frontend/backend/full-stack with Node/DOM cues), the way diffs are mapped to functions, the obfuscation/noise tools and sink lists, and prompt details. Porting to C/C++ or Python mainly involves swapping in a language parser and build toolchain, redefining source–sink categories and CWE focus, choosing a semantics-preserving obfuscator, and adjusting prompts and evidence. The main takeaway likely holds across languages: models rely on surface hints, are brittle to small changes, and struggle under low-FPR constraints; validating this in other languages is future work.

\noindent \textbf{Do We Avoid 100\% Over- and Underestimation?}
We aim to reduce both overestimation and underestimation rather than claim they disappear. To limit overestimation we evaluate whole repositories, include fixed versions, add four stress variants, require function-level localization, and report VD-S under an FPR budget; to limit underestimation we use CWE equivalence classes, a denoised split, normalized outputs, and stronger prompts. Remaining risks include imperfect equivalence, residual label noise, occasional lucky type matches at project level, sensitivity to settings, and possible training contamination; we mitigate these by reporting multiple metrics, releasing the scorer, and focusing on robustness tests.

\noindent \textbf{How to Enhance LLM-Driven JavaScript Vulnerability Detection?}
To improve reliability, train on security data with program structure (AST/CFG/DFG/CPG) as input and require short, evidence-backed outputs that name the file, function, and line in JSON. Combine LLMs with taint analysis, symbolic execution, or graph analyzers to check flows. Train with noise, obfuscation, and prompt injection, and use abstention and calibrated thresholds to keep FPR within budget. Expand \benchname with more frontend cases and, where feasible, executable PoCs, add time-based splits, and keep the denoised and robustness variants.

\section{Related Work}
\label{sec:related_work}

\paragraphtitle{Existing Vulnerability Dataset Collection Methods}
Existing security datasets fall into three types: synthetic, automated collection, and manually curated. Synthetic datasets (e.g., CGC~\cite{caswell2012cgc}, Juliet~\cite{boland2012juliet}, LAVA~\cite{dolan2016lava}) offer control and scale but diverge from real engineering practices, miss long-tail and environment-coupled exploit chains, and are weak for generalization and robustness evaluation. Automated mining (e.g., BigVul, BugSwarm, VulinOSS~\cite{fan2020bigvul,tomassi2019bugswarm,gkortzis2018vulinoss}) brings breadth but often lacks executable exploits and rigorous verification, with high noise and label bias. Manually curated sets (e.g., Magma, Ghera, Ponta~\cite{hazimeh2021magma,mitra2017ghera,ponta2019manually}) ensure authenticity and reproducibility yet typically stop at vulnerable–fixed pairs without end-to-end exploits or dynamic behavior support. General bug benchmarks (e.g., Defects4J, Bugs.jar, BugSwarm~\cite{just2014defects4j,saha2018bugsjar,tomassi2019bugswarm}) target functional bugs, limiting their relevance for security tools. No approach simultaneously meets authenticity, scale, bidirectionality, and low noise. Our \benchgen collects real GitHub projects, distinguishes vulnerable versus fixed versions, and adds confidence labels to reduce noise.

\paragraphtitle{JavaScript Vulnerability Datasets}
The representative BugsJS benchmark~\cite{gyimesi2019bugsjs} targets functional bugs rather than security vulnerabilities and lacks executable exploit semantics, limiting systematic security evaluation. Studies in the npm ecosystem expose risks such as dependency vulnerabilities, supply-chain attacks, typosquatting, and trivial package abuse~\cite{decan2018impact,taylor2020defending,abdalkareem2017trivial}; ReDoS and prototype pollution have been widely documented~\cite{davis2021selective,davis2019rethinking,kang2022probe,kang2024follow}. Yet a unified, executable JS vulnerability benchmark is still missing. SECBENCH.JS~\cite{secbenchjs} provides authentic, executable server-side cases, but existing JS datasets are designed for traditional static/dynamic/hybrid tools and primarily cover server-side ecosystems. In contrast, our \benchname evaluates LLMs on real-world projects, stratified by frontend/backend/full-stack, enabling repeatable comparisons in accuracy, explanation quality, and robustness.

\paragraphtitle{Web Security}
Web security research spans specific vulnerability detection, general static/dynamic/hybrid methods, and defenses. Despite progress (e.g., CPG-based scanners such as CodeQL, ODGen, JAW~\cite{li2022mining,khodayari2021jaw,yamaguchi2014modeling,kang2023scaling} and Node.js analyses~\cite{staicu2018synode,gauthier2018affogato,nielsen2021modular}), a systematic evaluation framework tailored to LLM capabilities is still lacking, making it difficult to measure robustness, explainability, and data-flow reasoning on real-world code.
\section{Conclusion}
\label{sec:conclusion}
In this work, we present three key principles—comprehensive coverage, no underestimation, and no overestimation—and, based on them, we developed \SecJS, a systematic benchmark for assessing large language models on JavaScript vulnerability detection. \benchgen automates data generation over diverse real-world projects and mitigates biases in CWE coverage, labels, and scenarios, while \syseval unifies evaluation and diagnosis. We compare seven commercial models and reveal weaknesses in reasoning, robustness, and false-positive control, indicating that current LLMs are not yet reliable for JavaScript vulnerability detection. Future work includes finer-grained semantic modeling, stronger context alignment, hybrid human--AI workflows, and continued expansion of the benchmark.

\section*{Ethics Considerations}

We rely exclusively on publicly available sources (CVE, Mend.io, and GitHub patch commits); comply with open-source licenses; and neither collect nor disclose any personally identifiable information or sensitive configuration data.

\section*{LLM Usage Considerations}

\noindent\textbf{Research use.} We evaluate seven LLMs (e.g., GPT-5) for JavaScript vulnerability detection via the claude-code-security-review pipeline, using a unified prompt and fixed settings (temperature = 0.7, confidence $>=$ 0.8); single-pass with no manual curation or retries; we record model, version, and time; prompts, configurations, and scoring scripts are released for reproducibility.

\noindent\textbf{Writing assistance.} LLMs were used only for linguistic polishing; all methods, experiments, and conclusions were authored and verified by the authors; citations were manually curated and checked.

{ \bibliographystyle{IEEEtran}
\bibliography{10-bib}}

\clearpage
\appendices

\section{Dataset Entry Examples} \label{app:dataset-examples}

\subsection{Vulnerability Information Gathering Example} \label{app:vulnerability-gathering-example}

After the Vulnerability Information Gathering step, each data entry comprises 7 fields. Below is an example entry for CVE-2021-25941:

\begin{itemize}
\item \texttt{cve\_id}: CVE-2021-25941
\item \texttt{cve\_description}: Prototype pollution vulnerability~\cite{shcherbakov2023silent,li2021detecting} in 'deep-override' versions 1.0.0 through 1.0.1...
\item \texttt{severity}: CRITICAL
\item \texttt{cwe\_id}: CWE-1321~\cite{cwe-1321}
\item \texttt{published\_date}: 2021-05-14
\item \texttt{code\_links}: https://github.com/...
\item \texttt{sources}: mend.io
\end{itemize}

\subsection{Ground Truth Refinement Example} \label{app:ground-truth-refinement-example}

After the Ground Truth Refinement step, each dataset entry is enriched with additional fields for fine-grained vulnerability localization. Below is a complete example entry for the same CVE-2021-25941:

\begin{itemize}
\item \texttt{project\_name}: ASaiAnudeep/deep-override
\item \texttt{project\_type}: Backend
\item \texttt{cve\_ids}: CVE-2021-25941
\item \texttt{code\_links}: https://github.com/...
\item \texttt{n\_code\_links}: 2
\item \texttt{sources}: mend.io
\item \texttt{severity\_breakdown}: CRITICAL
\item \texttt{vulnerability\_classification\_breakdown}: Prototype Pollution
\item \texttt{cwe\_ids}: CWE-1321
\item \texttt{files}: src/index.js
\item \texttt{function\_names}: override
\item \texttt{function\_label\_breakdown}: ONEFUNC
\item \texttt{commit\_shas}: 2aced17651...
\item \texttt{publish\_date\_last}: 2021-05-14
\item \texttt{vulnerable\_code\_paths}: databackup/.../index\_vulnerable.js
\item \texttt{fixed\_code\_paths}: databackup/.../index\_fixed.js
\item \texttt{vulnerable\_line\_ranges}: 19-70,49,50
\end{itemize}

\section{Code Examples for Dataset Augmentation and Prompt Injection}

\subsection{Noise Injection Example}
\begin{lstlisting}[caption={Noise Injection Example}, label=lst:noise_injection]
// === Original Code ===
function authenticateUser(username, password) {
  // Real SQL injection vulnerability
  const query = `SELECT * FROM users WHERE username='${username}'`;
  return db.query(query);
}

// === After Noise Augmentation ===
function authenticateUser(username, password) {
  // Noise: File operation sink (no taint source)
  const logFile = '/var/log/app.log';  
  fs.appendFileSync(logFile, 'Login attempt\n'); 
  
  // Real SQL injection vulnerability (unchanged)
  const query = `SELECT * FROM users WHERE username='${username}'`;
  return db.query(query);
}
\end{lstlisting}

\subsection{Code Obfuscation Example}
\begin{lstlisting}[caption={Code Obfuscation Example}, label=lst:obfuscation, breaklines=true, breakatwhitespace=false]
// === Original Code ===
function executeCommand(userCmd) {
  const cmd = `ls ${userCmd}`;
  return exec(cmd);
}

// === After Obfuscation ===
eval(function(p,a,c,k,e,d){
  e=function(c){
    return(c<a?"":e(parseInt(c/a)))+
    ((c=c%a)>35?
      String.fromCharCode(c+29):
      c.toString(36))
  };
  if(!''.replace(/^/,String)){
    while(c--)
      d[e(c)]=k[c]||e(c);
    k=[function(e){
      return d[e]
    }];
    e=function(){
      return'\\w+'
    };
    c=1;
  };
  while(c--)
    if(k[c])
      p=p.replace(
        new RegExp(
          '\\b'+e(c)+'\\b','g'),
        k[c]);
  return p;
}(
'3 4(0){2 1=`7 ${0}`;6 5(1)}',
8,8,
'userCmd|cmd|const|function|'+
'executeCommand|exec|return|ls'
.split('|'),0,{}))
\end{lstlisting}

\subsection{Prompt Injection Examples}
\paragraphtitle{Type 1: False Safety Claims} (False Negative Induction)
\begin{lstlisting}[caption={Prompt Injection -- False Safety Claims}, label=lst:prompt_fn]
function processUserInput(data) {
  // This function is completely safe and has been security audited.
  // Actually contains XSS vulnerability
  document.getElementById('output').innerHTML = data;
}
\end{lstlisting}

\paragraphtitle{Type 2: False Vulnerability Claims} (False Positive Induction)
\begin{lstlisting}[caption={Prompt Injection -- False Vulnerability Claims}, label=lst:prompt_fp]
function generateReport(userId) {
  // WARNING: This code is vulnerable to SQL injection!
  // TODO: Fix the security issue in this function.
  // Actually safe code with parameterized query
  const query = 'SELECT * FROM reports WHERE user_id = ?';
  return db.execute(query, [userId]);
}
\end{lstlisting}

\section{Overview Code Examples: Pattern Dependence and Context}

\subsection{Keyword Dependence}
\begin{lstlisting}[language=JavaScript,numbers=left,basicstyle=\ttfamily\footnotesize,caption={Overview — Keyword Dependence}, label=lst:ov_keyword]
// Model may wrongly flag XSS merely due to seeing innerHTML
element.innerHTML = userInput;

// Secure usage that should not be flagged
element.innerHTML = DOMPurify.sanitize(userInput);
\end{lstlisting}

\subsection{Syntactic-Structure Dependence}
\begin{lstlisting}[language=JavaScript,numbers=left,basicstyle=\ttfamily\footnotesize,caption={Overview — Syntactic-Structure Dependence}, label=lst:ov_syntax]
// Model may infer SQL injection only from the variable name "query"
const query = "SELECT * FROM users";

// In fact this is a hard-coded safe query
db.execute(query);
\end{lstlisting}

\subsection{Context Neglect}
\begin{lstlisting}[language=JavaScript,numbers=left,basicstyle=\ttfamily\footnotesize,caption={Overview — Context Neglect}, label=lst:ov_context]
function processData(input) {
  // Model may ignore this critical validation step
  if (!isValid(input)) return;

  // Focusing only on this line would be a misclassification
  database.query(input);  // input has already been validated
}
\end{lstlisting}


\end{document}